%% file: bead2.tex
\documentstyle[epsfig]{jfm} 

\title[The beads-on-string structure of viscoelastic threads]
{The beads-on-string structure of viscoelastic threads}

\author[Ch. Clasen et al.]
{Christian Clasen$^{1,5}$, Jens Eggers$^{2,6}$, Marco A. Fontelos$^3$, 
Jie Li$^4$, and Gareth H. McKinley$^1$
  }
\affiliation{$^1$Hatsopoulos Microfluids Lab, 
Department of Mechanical Engineering, MIT, Cambridge, Mass. 02139, USA \\
$^2$Universit\"at Gesamthochschule Essen, Fachbereich Physik,
45117 Essen, Germany \\
$^3$ Departamento de Ciencia e Ingenieria,
Universidad Rey Juan Carlos,
C/ Tulip\'an S/N, 28933 M\'ostoles, Madrid, Spain. \\
$^4$BP Institute \& Engineering Department,
University of Cambridge, Madingley Road
Cambridge, CB3 0EZ, United Kingdom \\ 
$^5$present address: Institut f\"ur Technische und Makromolekulare Chemie,
Bundesstr. 45, 20146 Hamburg, Germany \\
$^6$present address: School of Mathematics, 
University of Bristol, University Walk, \\
Bristol BS8 1TW, United Kingdom 
           }
\pubyear{2002}
\date{?? and in revised form ??}
\begin{document}

\maketitle

\begin{abstract}
By adding minute concentrations of a high molecular weight polymer, liquid
jets or bridges collapsing under the action of surface tension develop a
characteristic shape of uniform threads connecting spherical fluid drops. 
In this paper, high-precision measurements of this beads-on-string
structure are combined with a theoretical analysis of the limiting case
of large polymer relaxation times, for which the
evolution can be divided into two distinct regimes. This excludes the very
late stages of the evolution, for which the polymers have become fully
stretched. For times smaller than the polymer relaxation time, over which
the beads-on-string structure develops, we give a simplified local
description, which still contains the full complexity of the problem. At
times much larger than the relaxation time, we show that the solution
consists of exponentially thinning threads connecting almost spherical
drops. Both experiment and theoretical analysis of a one-dimensional 
model equation reveal a self-similar structure of the corner where 
a thread is attached to the neighbouring drops. 
\end{abstract}

\input{introduction2}
\input{simulation2}

\input{localapprox2}

\input{expon2}
\input{conclusion2}
\begin{acknowledgments}
We are grateful to the Deutsche Forschungsgemeinschaft, who 
funded M. F. and J. L.'s visit to Essen through grant SFB237. G.H.M. and C.C.
would like to thank the Dupont-MIT Alliance and the
NSF-MRI program for supporting the experimental portion of this work.
Daniel Bonn contributed a careful reading of the manuscript and
very useful discussions.
\end{acknowledgments}        
\input{appendix}

\input{beadbib}
\end{document}

%% file: introduction2.tex
\section{Introduction}

Understanding the behaviour of polymeric free-surface flows is of enormous
importance for a wide variety of applications in the chemical
processing, food and consumer products industries. Operations
such as ink jet printing, spraying of fertilizers, paint-leveling,
misting, bottle-filling and roll coating
are all controlled by interactions between the  non-Newtonian stresses
in the bulk and capillary stresses at the deformable free-surface.
  Long-chained macromolecules are also
ubiquitous in biological fluids, and very significantly affect the
corresponding free-surface
dynamics. If one places a small drop of saliva between two fingers
and pulls them apart, the resulting liquid bridge does not collapse,
but an extremely fine thread remains for several seconds. The lifetime of this
bridge is intimately connected with the molecular weight and conformation
of the proteins (mucins) and hormones in the saliva. Measurement of 
such filament lifetimes
in biofluids such as mucus or saliva can be used as a fertility indicator
\cite{KK79}.

A number of recent studies have promulgated the idea of using the 
capillarity-induced
thinning of a liquid filament as a rheometric device for quantifying the
properties of complex fluids in predominantly extensional flows
(\cite{Baz90,Stel00,Trip00}). A typical configuration is
shown in Figure 1. A liquid bridge of the fluid to be tested 
is initially formed
between two coaxial cylindrical plates and a step uniaxial strain is 
then imposed on
the bridge to extend it beyond the static (Plateau) stability limit.
The liquid filament then undergoes a capillary-thinning process towards
a final breakup event. The no-slip
boundary condition at the endplates retards the radial flow near the end plates
and thus imposes a well-defined initial axial perturbation or 'neck' on the
liquid column which controls the location of the subsequent necking 
process.  The
  time evolution in the local filament radius is monitored optically 
using either
a laser micrometer or high-speed video imaging.  To convert such measurements
of filament evolution into an extensional viscosity or other material function
characterizing free-surface extensional flows of complex fluids, it 
is necessary
to understand the balance of forces acting on the fluid filament.
Capillary pressure drives the thinning process whilst viscous and 
elastic stresses (plus
gravity) resist the necking of the fluid thread.   The slenderness of the fluid thread
induced by the step strain means that a one-dimensional approximation to the
equations can be useful and the large viscosity of many polymeric fluid
systems means that inertial effects can often be neglected over the majority of
the capillary-thinning process. \cite{EH97} provide a detailed
discussion of the evolution of a cylindrical thread of viscoelastic 
fluid undergoing
capillary-driven breakup.  A central result of both this work and 
earlier studies
is that there can be a lengthy intermediate regime in which inertial, 
viscous and
gravity forces are all negligible and  elastic and capillary forces 
balance each.
In this regime the local extension rate in the neck is constant and
the radius of the filament decreases exponentially in time. Measurement of this
rate of thinning enables a direct determination of the characteristic 
relaxation time
of the viscoelastic fluid. Such observations have been
found to be in quantitative agreement with data obtained in 
extensional rheometers (\cite{AM01}).

\begin{figure}
\begin{center} 
\psfig{figure=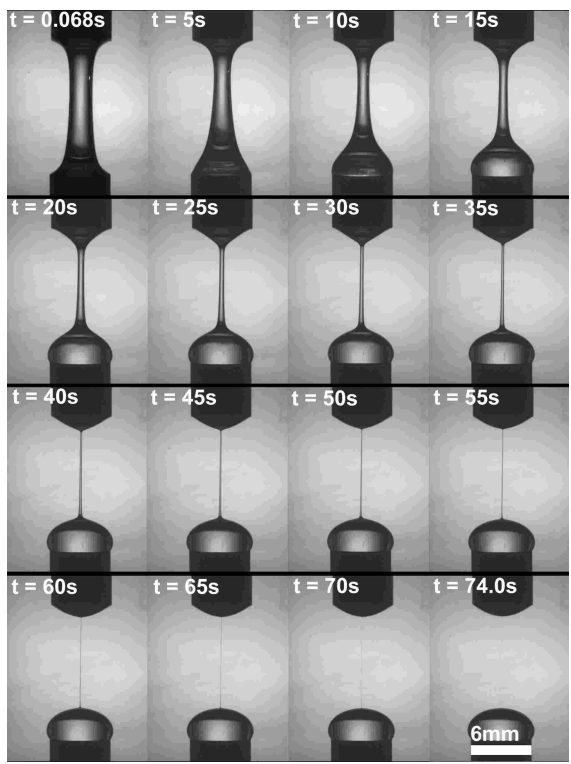,height=10cm} 
\end{center}
\caption{
Experimental images of a collapsing liquid bridge of polymer solution
in a viscous solvent (plate radius $R_0 = 3 mm$, distance 13.8 $mm$). 
The surface tension is $\gamma=37mN/m$, 
the density $\rho=1026kg/m^3$. The solvent and polymeric contributions 
to the viscosity are $\eta_s=65.2 Pa s$ and $\eta_p=9.8 Pa s$, 
respectively, the polymer 
timescale is $\lambda = 8.1 s$. Relative to the capillary timescale
$\tau=\sqrt{\rho R_0^3/\gamma}$ this results in a Deborah number 
of $De = 296$. 
    }
\label{clasen1}
\end{figure}

This local elasto-capillary balance has also been observed
in other experimental configurations including breakup of forced polymeric
jets (\cite{CW02}) and gravity-driven drop formation in viscoelastic polymer
solutions (\cite{ABMK01,Coop02}). In these  experiments, the fluids typically
have low viscosities and the initial stages of the process exhibit 
characteristics
of a self-similar solution resulting from a balance between capillary 
pressure and fluid
inertia (\cite{DHL98}) including a power-law decrease in the minimum 
radius. The rapid
growth in viscoelastic stresses within the fluid neck leads to a cross-over
to an exponential decrease in the minimum filament radius with time 
which is a hallmark
of the elasto-capillary balance.  However additional details of the 
self-similar solution (such as the spatial profile of the filament)
 are presently unknown.  By contrast,
in the absence of polymeric additives, the cross-over to a 
self-similar  'universal solution'
balancing viscosity, inertia and capillarity are now well-explored 
(\cite{E97, RRR01,CNB02}).

\begin{figure}
\begin{center} 
\psfig{figure=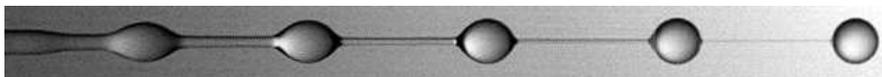,height=1cm}
\end{center}
\caption{
High speed videoimage of a jet of dilute (0.01 wt\%) aqueous 
polyacryamide solution (surface tension $\gamma = 62 mN/m$) undergoing 
capillary thinning. 
The sharp-edged jet orifice is at the left of the image 
(radius $R_0$ = 0.30 mm) and the free jet velocity is  30 cm/s.
The polymeric contribution to the viscosity is 
$\eta_p = 0.0119 Pa s$, and the polymer timescale is found to be 
$\lambda=0.012 s$.
This corresponds to a Deborah number of $De=18.2$.
  }
\label{clasen2new}
\end{figure}

  On close examination of a thinning viscoelastic jet,  a
string of tiny droplets can often be distinguished. This
'beads on a string' phenomenon was first described in \cite{GYPS69},
and has been reproduced in numerical simulations by \cite{BKMD86}.
 A representative image of
this viscoelastic jet break-up process is shown in \ref{clasen2new}. 
The jet consists of a series of cylindrical ligaments connecting
spherical beads. As the jet is convected from left to right, fluid
is forced by capillarity from the thinning ligaments into the
spherical droplets
% Jens - add commands to display fig 2 near here...

Most analytical studies of this structure have been performed using
simplifying assumptions about the slenderness of the liquid jet, see \cite
{Y93} for a review. However, despite a considerable number of studies
(see e.g.  \cite {GG82,EY84,BKMD86,FW90,SDK91,L92,R94,R95,CDK99})
a full analytical description of the beads-on-string phenomenon
is still open, even in the context of one-dimensional models.

In this paper we seek a self-similar solution 
that encompasses the elasto-capillary balance
documented in experimental observations in liquid bridges, pinching 
drops and thinning jets.
We  follow the spirit
of some of the earlier work by employing two simplifying assumptions:

First, we will consider the  simplest canonical model for a dilute
polymer solution, the so-called Oldroyd-B model (e.g. \cite{BAH87}). 
This model is most
easily derived from kinetic theory by treating a dilute solution of
polymer chains  as a suspension of non-interacting dumbbells, i.e.
as two  point masses or `beads' at which the viscous drag from the 
solvent is imparted,
connected by a Hookean spring. The beads are assumed to be convected 
by the flow without slip,
so in the extensional flow of a pinching thread, the beads are pulled 
apart, stretching the
spring. In return, the  dumbbell imparts a stress on the fluid, which 
results in a polymeric
contribution to the stress tensor which is proportional to the force 
in the spring. The
simplifying assumptions of a Hookean dumbbell are that the polymer chain 
exhibits  a
single relaxation time $\lambda$, and that the spring is infinitely
stretchable. The one-dimensional analysis of \cite{EH97} considered 
the more general
case of a spectrum of relaxation times (corresponding to a 
non-interacting suspension
of dumbbells with different spring constants). Their analysis showed that the
rate of stretching in the liquid thread is governed by the dumbbells with
the longest time constant and that all other modes relax and do not 
contribute to the
dominant balance, so this approximation of a single relaxation time is not
considered to be too limiting. Furthermore, a
number of recent experimental studies (e.g.
\cite{SAM96,ABMK01,AM01}) have utilized model dilute polymer 
solutions which are indeed
very well described by a single time scale over a wide range of extensions.
The additional assumption of infinite extensibility is bound to break 
down even for very long
polymers as the trajectories of the two beads diverge exponentially 
in an extensional flow.
In fact, it has been shown by \cite{R94} for the model to be treated in this
paper (and neglecting inertia) that a thread can {\it never} break up in
finite time. In our
study, we therefore disregard the final stages of breakup where the finite
length of the polymers begins to affect the necking process. This regime
has been considered by \cite{EH97}, and the filament radius ultimately
decreases linearly in time. 

The second simplifying assumption is that we are treating the flow inside
the fluid thread as effectively one-dimensional (e.g. \cite{FW90}). This is
consistent as long as the shape of the liquid column remains slender, i.e.
the characteristic radial variations are small compared to the variation in the axial
direction. This assumption is problematic near the ends of the fluid drops
in the beads-on-string structure. Following \cite{ED94}, we hope to at least
partially deal with this problem by keeping the {\it full} expression for
the mean curvature in the Laplace pressure, which drives the breakup. This
makes spherical drops exact static solutions of the equations, and ensures
that at least the surface tension terms are correctly accounted for.

We are left with a model that treats the liquid column as a set of
one-dimensional continuum equations for the fluid flow coupled with
equations describing the state of stress of the polymer chains in 
solution. A typical
experimental situation would be that of a jet ejected from a nozzle, or a
liquid bridge held between two circular end-plates. In all of the following,
we will choose the initial bridge or jet radius $R_0$
as a unit of length, and the
corresponding capillary time $\tau = \left(\rho R_0^{3}/\gamma \right)^{1/2}$
as unit of time, where $\gamma$ is the surface tension and $\rho$ the
density of the fluid. If $R_0 \approx $ 1mm, $\tau$ is about 4 ms for a
water-based solvent. Note that for a high viscosity fluid (also treated
in this paper) other time scales such as the viscous scale 
$\tau_{\eta} = \eta R_0/\gamma$ arising from a balance of
surface tension and viscosity, might be more appropriate. However,
to avoid confusion we will consistently use the inertial-capillary
time scale. There still remain three independent dimensionless
parameters in the problem. The time scale $\lambda$ of the polymer is
conventionally called a Deborah number, $De$, when made 
dimensionless with the
characteristic time scale of the system. In the present study we
thus have $De = \lambda/\tau$. Note that the Deborah number is 
`intrinsic' to
the fluid thread because it is defined
entirely in terms of material and geometric parameters. It does not contain the
rate of stretching in the fluid, since the flow is not forced but is
free to select its own rate of stretching, which may be spatially and/or
temporally inhomogeneous.
   The other two dimensionless parameters represent the relative 
contributions of viscous
stresses from the solvent and the polymer. There are a number of 
possible representations for
these parameters. The total dynamical viscosity for a dilute polymer solution
characterized by the Oldroyd-B model is given by
$\eta_0 =\eta_s + \eta_p$ and the relative importance of viscous effects
can thus be characterized
by the Ohnesorge number $Oh = \eta_0/\sqrt{\rho \gamma R_0}$ and the
solvent viscosity ratio $S = \eta_s/\eta_0$. An alternative representation
which we use below is to separate the relative dimensionless 
contributions of the
kinematic viscosity $\nu_s = OhS$ of the solvent, and the polymeric 
contribution to the
viscosity $\nu_p = Oh(1-S)$. All these
material  constants have been made dimensionless using $R_0$, 
$\gamma$ and $\tau$
as the characteristic scales.

Even if one only considers the case of a fixed initial condition, and
neglects gravity (say in a liquid bridge configuration), the problem is still
one of daunting complexity. In particular, the behaviour of the liquid bridge
need not be spatially uniform throughout: in some regions the rate $v^{\prime}$ at
which fluid elements are stretched may be smaller than $1/(2 De)$. 
In this case
polymer relaxation overcomes flow stretching, and the polymers remain
coiled, leading to Newtonian-like behaviour. In other parts of the 
flow stretching
overpowers relaxation, and the polymers become stretched in unison with the
exponentially diverging paths of fluid elements. In these regions, the
polymeric contribution to the stress grows accordingly, and threads form.
The transition from a flow regime with $2 v^{\prime}De < 1$ to one with 
$2v^{\prime}De > 1$, where exponential stretching of polymers may occur, has
been called the ``coil-stretch'' transition by \cite{dG74}.

It is very difficult
to describe this complexity in full generality, so in the
following we will restrict ourselves to the case of large $De \gg 1$,
implying that the non-Newtonian polymer contribution is significant at all times.
Physically this means that $De$ is much larger than the initial 
time scale of the liquid bridge's evolution, which is set by the linear stability of
the fluid thread. At low
viscosities, $Oh = \nu_s + \nu_p < 1$, this time scale is $O(1)$ by virtue of
the chosen time scale for non-dimensionalization.
By contrast, for fluids with large viscosities it is set by $\tau_{\eta} = \tau/Oh$, 
and we thus require $De/Oh \gg 1$.
Note that with the present scaling, $1/Oh^2$ is the ratio of the external length
scale $R_0$ and the intrinsic scale of the fluid, $\ell_{\nu} =
\nu^2/(\gamma\rho)$.

%The parameter $1/\nu$ is often called a Reynolds number
%(e.g. \cite{E97}), or alternatively an Ohnesorge number (e.g. \cite{O36}).

\begin{figure}
\begin{center} 
\psfig{figure=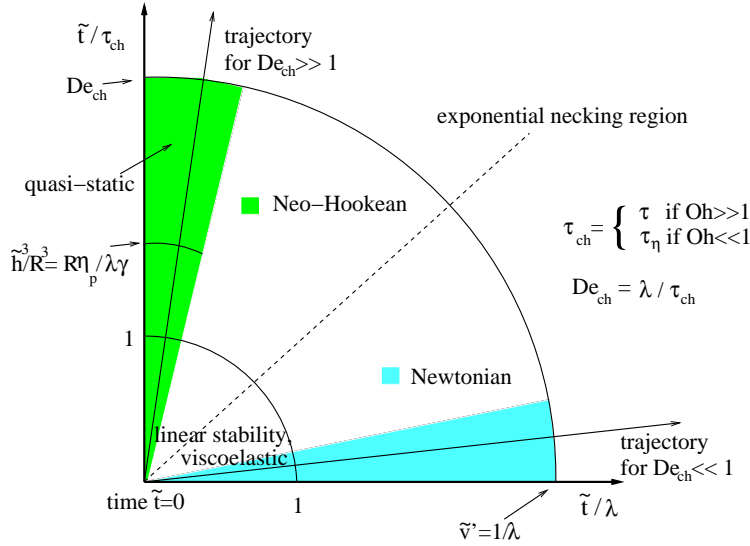,height=13cm,angle=-90} 
\end{center}
\caption{
Schematic diagram showing the transient evolution of viscoelastic 
fluid threads.
Trajectories correspond to rays of constant slope, given by the 
the `intrinsic' Deborah number
$De = \lambda/\tau_{ch}$, where $\tau_{ch}$ is the characteristic 
capillary timescale (depending on whether viscous or inertial 
effects are more important in the bulk). The focus of this paper is 
on the sector corresponding to large slopes, and more details are given 
in the text. For small slopes the behaviour is Newtonian for a significant
portion of the evolution, until the stretch rate $\tilde{v}'$ exceeds
the inverse polymer relaxation time. 
}
\label{clasen3new}
\end{figure}

With these assumptions, we can divide the transient evolution of the jet into distinct
stages as shown schematically in Figure \ref{clasen3new}. The dimensional time $\tilde{t}$
elapsed is scaled with both the polymer relaxation time $\lambda$ and the
appropriate time scale for evolution of the fluid thread $\tau_{ch}$
(which depends on the value of the Ohnesorge number).
Fluid elements in the jet evolve along pathlines of constant slope 
$\lambda/\tau_{ch} = De_{ch}$. In each regime,  we can provide a simplified 
description by balancing the dominant terms in the governing equations.
The focus of this paper will be on the sector characterized by large 
$De_{ch}$.

In the early elastic time regime, $t \ll 1 \ll De$, there is no significant 
decay of polymer stretching. The fluid thus responds as a neo-Hookean 
elastic material.
This allows the effect of the polymers to be written as a {\it
local} contribution to the pressure, given in terms of the interface shape
(i.e. the local accumulated strain) alone.
The parameter determining the magnitude of this contribution is 
the dimensionless elastic modulus of the material
$G = \nu_p/De$, which is (up to universal constants) proportional to the
polymer {\it concentration}. Depending on the viscosity, the dynamics of the
bridge can be quite complex. In particular, for low viscosities, capillary
waves can travel along threads and rebound off drops (\cite{FL03}). 
Threads are also shown to support elastic waves. 

For $De \gg t \gg 1$, as polymers become sufficiently stretched to
counter surface tension forces, the simplified, local system of equations
converges to a {\it stationary} solution, maintained by the stress in the
polymers with no possibility of relaxation. This stationary solution,
originally found by \cite{EY84}, already exhibits the beads-on-string
structure, but with a thread of radius $h_{thread}=(G/2)^{1/3}$ 
to be computed in section \ref{sta}.
The transition of the initial evolution 
to the region marked ``quasi-static'' in Fig. \ref{clasen3new}
thus occurs approximately when this radius is reached. The name 
``quasi-static'' refers to the fact that the solution can only be 
regarded as stationary on time scales much smaller than the polymer 
relaxation time. 

Indeed, to proceed beyond this stage one has to take the viscoelastic 
relaxation of the polymer chains into account. 
The structure of the solution is that of cylindrical
filaments which thin at an exponential rate $\exp(-t/3De)$ as a result of
the local balance between
elasticity and capillarity. The  filaments connect an
arbitrary distribution of droplets, which approach a static, spherical shape.
A similarity solution describes the crossover between the cylindrical 
thread and the
neighbouring droplet. Towards the thread, the solution asymptotes towards a
constant thickness, in the direction of the drop it fits onto the spherical
shape of the drop. Since the asymptotic shape is that of threads of
vanishing radius connecting perfect circles, a corner develops, with the
interface slope diverging on the side of the drop. Thus the description of
this part of the flow is beyond the capabilities of a slender-jet model.
However, by keeping the full curvature term, we still hope to obtain a
reasonable description of this transition region. 
At low values of $De$, the initial
stages of the necking process at short times $t < De < 1$ 
are controlled by the linear stability
of a viscoelastic fluid thread (\cite{Mid65,FJ03}); 
however ultimately at longer times $De < 1 < t$ there will 
still be a crossover to the exponential necking regime, 
provided the polymer chains are sufficiently extensible. 
We do not consider these regimes further in the present study.

Our paper is organised as follows: In the next section we develop and
motivate the lubrication equations to be used for the remainder of this
paper. A numerical simulation illustrates the regimes to be analysed
below. The third section is devoted to the study of the neo-Hookean 
regime, where polymer relaxation can be neglected. 
First we derive a simplified, local description from the model equations.
This local description is then used to compute the asymptotic thread radius,
and to describe the propagation of elastic waves on the thread. The
fourth section deals with the long time regime at finite $De$ 
for which exponential
thinning of threads is observed. After giving a qualitative description of the
shape and flow inside the thread, we introduce a similarity description
valid in the corner where a cylindrical thread meets a spherical 
drop. If $De$ is large
enough to make only elastic and surface contributions relevant, we can compute
all but one of the free parameters of the solution. This last parameter,
the thread radius, can be estimated by matching to the early-time regime.
The numerical results are compared with experimental observations using
a dilute solution of monodisperse polystyrene which is well-described by the
Oldroyd-B constitutive model. Measurements of the evolution in the mid-filament
radius and the evolution of the spatial profile of the filament are
well-described by the theory.
In the final section we discuss work that remains to be done within
the framework of the present model, as well as perspectives for
inclusion of other effects that lie beyond it.

%% file: simulation2.tex
\section{Model and simulation}
\subsection{One-dimensional equations}
In this paper we confine ourselves to the study of a simplified version of
the Oldroyd B model for polymeric liquids, assuming that the
radius $h(z,t)$ of the liquid column varies slowly. Thus the variation of
hydrodynamic variables inside the column is also small, and we can confine
ourselves to the leading order approximation in an expansion in the radius.
For example, $v(z,t)$ below is the axial velocity at the centre of the jet.
A derivation of the relevant equations has been given in \cite{FW90}, so we
just give the final result and briefly discuss its physical significance:

\begin{eqnarray}
&& \frac{\partial h^2}{\partial t}+ \frac{\partial}{\partial z}%
\left(vh^2\right) = 0  \label{b1} \\
&& \frac{\partial v}{\partial t}+v\frac{\partial v}{\partial z}= -\frac{%
\partial \kappa }{\partial z}+ 3\nu_s\frac{1}{h^{2}}\frac{\partial }{%
\partial z} \left(h^{2}\frac{\partial v}{\partial z}\right) + \frac{1}{h^{2}}%
\frac{\partial }{\partial z} \left(h^{2}(\sigma _{zz}-\sigma_{rr})\right)
\label{b2} \\
&& \frac{\partial \sigma _{zz}}{\partial t} + v\frac{\sigma _{zz}}{\partial z%
} = 2\frac{\partial v}{\partial z}\sigma_{zz} + 2\frac{\nu _{p}}{De}\frac{%
\partial v}{\partial z} - \frac{\sigma _{zz}}{De}  \label{b3} \\
&& \frac{\partial \sigma _{rr}}{\partial t} + v\frac{\sigma _{rr}}{\partial z%
} = -\frac{\partial v}{\partial z}\sigma_{rr} - \frac{\nu _{p}}{De}\frac{%
\partial v}{\partial z} - \frac{\sigma _{rr}}{De}  \label{b4}
\end{eqnarray}

Equation (\ref{b1}) expresses volume conservation, (\ref{b2}) is the
momentum balance equation in the one-dimensional approximation. The first
term on the right of (\ref{b2}) is the gradient of the Laplace pressure,
which is the main driving force. With the present scaling, the 
capillary pressure gradient
is simply the gradient of the mean curvature $\kappa$, 
which for later convenience
will also be written as
\begin{equation}
-\frac{\partial \kappa }{\partial z}=\frac{1}{h^{2}}\frac{\partial }{%
\partial z}\left( h^{2}\left( \frac{h_{zz}}{(1+h_{z}^{2})^{\frac{3}{2}}}+%
\frac{1}{h(1+h_{z}^{2})^{\frac{1}{2}}}\right) \right) \equiv \frac{1}{h^{2}}%
\frac{\partial }{\partial z}\left( h^{2}K\{h\}\right) \;.  \label{ber1}
\end{equation}

The second term on the right of (\ref{b2}) is the Newtonian contribution to
the viscosity, multiplied by the dimensionless viscosity $\nu_s$ of 
the solvent.
Finally the last term is the polymeric contribution; $\sigma_{zz}$ and 
$\sigma_{rr}$ are the diagonal terms of the extra stress tensor. The other
components of the polymeric stress tensor do not enter at leading order.
Using (\ref{ber1}), equation (\ref{b2}) can finally be rewritten such that
the inertial terms on the left are balanced by gradients of the tension in
the thread:
\begin{equation}  \label{tension}
\frac{\partial v}{\partial t}+v\frac{\partial v}{\partial z}= \frac{1}{h^{2}}%
\frac{\partial}{\partial z}\left[ h^{2}\left(K + 3\nu_s\frac{\partial v}{%
\partial z} + \sigma _{zz}-\sigma_{rr}\right)\right]\quad.
\end{equation}

For very viscous fluids the inertial terms on the left are negligible, and
the terms in parentheses equal a (generally time-dependent) constant. If
velocity gradients $\partial v/\partial z$ are small,
(\ref{b3}) and (\ref{b4}) describe an additional {\it Newtonian}
contribution $\nu_p$ to the total steady-state shear viscosity $\nu = \nu_s + \nu_p$. 
The presence of polymers also results in fluid viscoelasticity and the polymeric
stress in the fluid responds with a dimensionless
time delay of $De$, which is
represented by the relaxation terms $\sigma_{ij}/De$ on the right
of (\ref{b3}) and (\ref{b4}). Finally a
crucial term for the physics of the following is the first term on the right
of (\ref{b3}) and (\ref{b4}), which describes the interaction of the polymer
with the flow. In an extensional flow $\partial v/\partial z$ is positive,
so the stress in the axial direction grows as the dumbbells modeling the
polymeric contribution to the stress are stretched, while
it decays in the radial direction.

\subsection{Beads on a string}
We are now in a position to study the behaviour of the model 
for various initial conditions and to compare to experiment. 
First, we simulate the decay of a long, initially unstretched cylinder 
of fluid. 
For our simulations we have used a numerical code analogous to the one
developed earlier by \cite{ED94,E97} for Newtonian flows. It is fully implicit
and uses adaptive regridding of the numerical mesh to fully
resolve the fine structure of the flow. This is crucial to be able to
describe some of the last stages of thread formation to be investigated
in detail in section 4. We found that the demands on the solution of the
implicit equations are much greater than in the Newtonian case, owing to a
larger range of time scales in the flow. In general, several iterations of a
Newton scheme were necessary for convergence, and significant restrictions
had to be put on the time step.
To further test for possible problems inherent in our numerical scheme,
Another {\it explicit} code was also developed independently. In addition,
all fields were represented in a uniform grid, as opposed to the staggered
grid of the implicit code. The explicit code performed quite well except for
the highest viscosities, where the time step imposed by the Courant condition
became prohibitively small. In all our tests, we found no significant
disagreement between the results of the two codes, so below we will no
longer specify which one was used to obtain a specific result.

\begin{figure}
\begin{center}
\psfig{figure=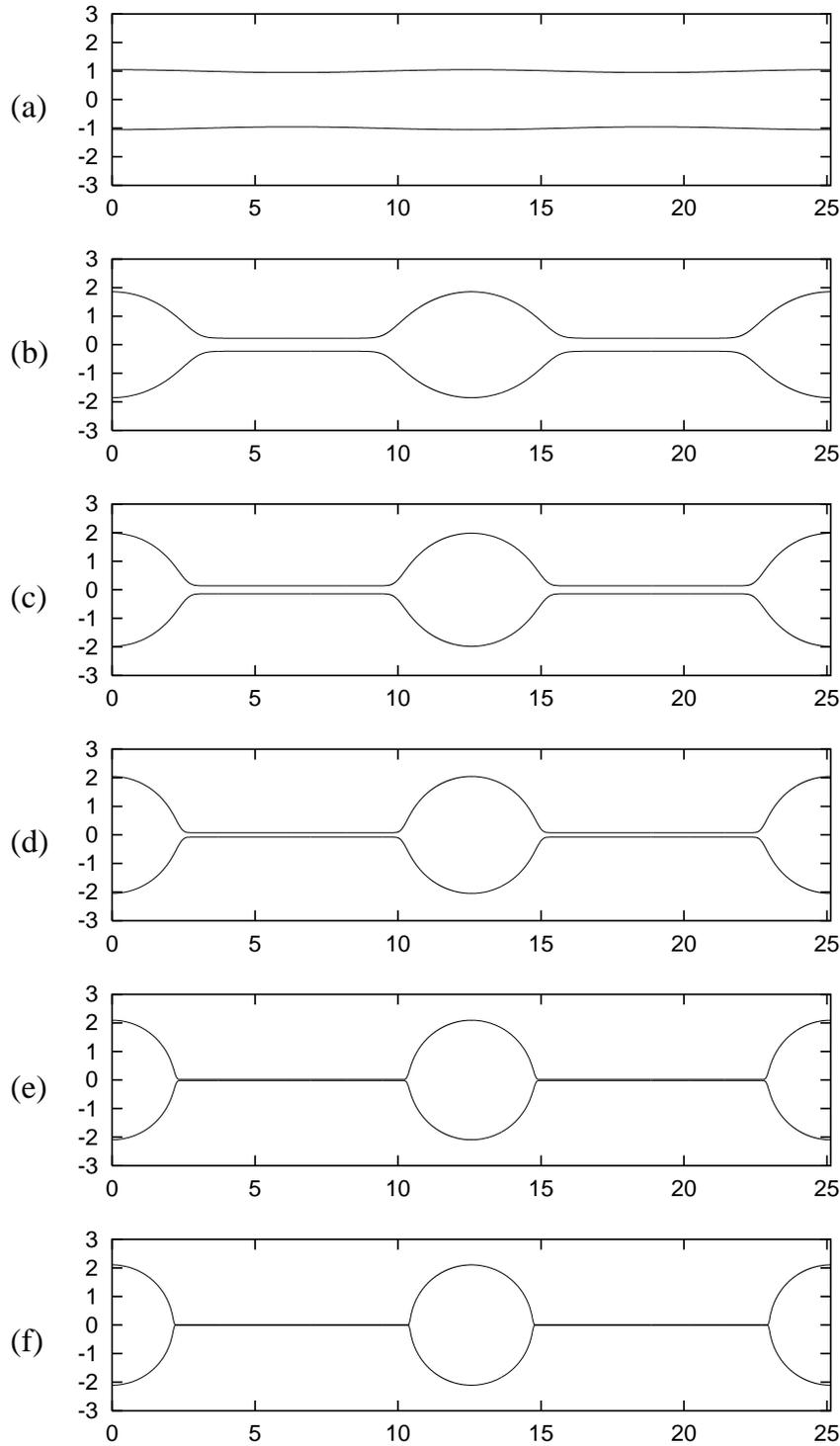}

\end{center}
\caption{
A typical series of profiles with periodic boundary conditions
and period $L=4\pi$. The dimensionless parameters
are $\nu_s=0.79$, $\nu_p=2.37$, and $De = 94.9$.
A very small sinusoidal
perturbation was added in order to make the filament collapse.
After the rapid formation of the beads-on-string structure,
one observes the slow thinning of the thread. The relative dimensionless
times of each profile
are (a) 0.0, (b) 31.6, (c) 158.1, (d) 316.2, (e) 632.5 and (f) 948.7.
}
\label{fig1}
\end{figure}

\begin{figure}
\begin{center}
\psfig{figure=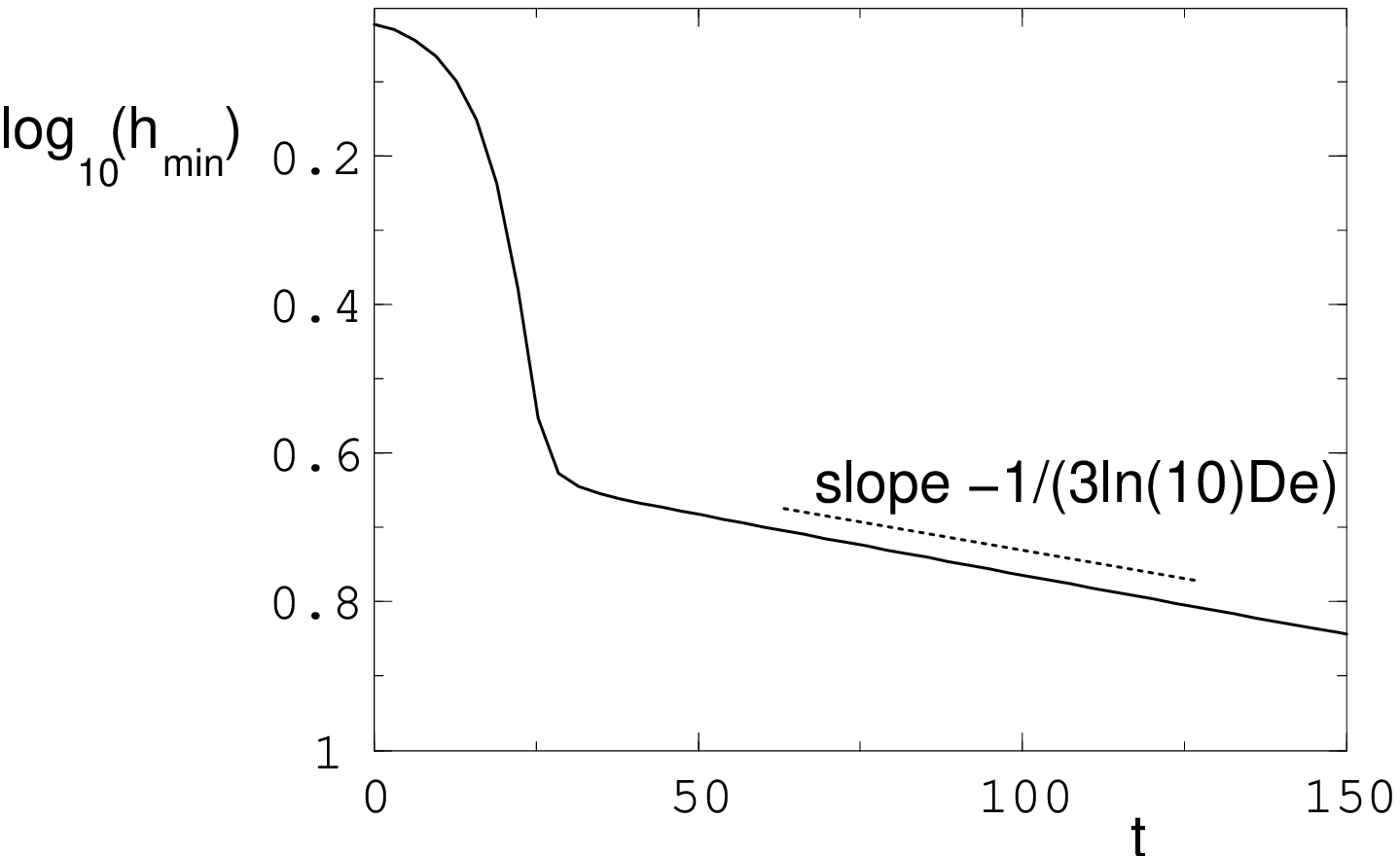}
\end{center}
\caption{
The minimum radius corresponding to the profiles shown in Fig.\ref{fig1}.
One clearly observes a rapid initial motion for times small compared
to $De = 94.9$, followed by an exponential thinning at rate $1/(3De)$.
}
\label{fig2}
\end{figure}

% XX Jens: I think that times were rescaled by a factor of De or something,
% so these long dimensionless times are wrong i think????

Figures \ref{fig1} and \ref{fig2} give an idea of the typical
behaviour of a liquid filament in the absence of gravity, described
by (\ref{b1}-\ref{b4}). The parameters were chosen to be identical to
those of Fig. 2 in \cite{CDK99}; that is in the present scaling, $\nu_s=0.79$,
$\nu_p=2.37$, and $De = 94.9$,
except that our initial perturbation is smaller.  The periodic boundary
condition is used here with the period $L = 4 \pi$.
The relaxation time scale for the polymer solute ($\lambda$) is 
approximately 95 times the capillary
time scale, so at short times, viscoelastic relaxation effects
can be neglected. Additionally, the Ohnesorge number is
$Oh = \nu_s + \nu_p = 3.16$, so the flow is significantly damped by viscosity.
Fig. \ref{fig1} shows a sequence of profiles for a length of two periods
($8 \pi$) at different times. After
an initial period described by linear theory, a thin thread is formed
between two round drops. The filament necks in the central region and
the fluid expelled from this neck is accumulated in the two beads.
The   evolution in the minimum radius of the thread with
time is shown in  Fig. \ref{fig2}. There is an initial sharp decrease,
until a sudden change in the
slope occurs. This happens when the elastic stresses which build
up in the deforming liquid bridge become dominant and a local elasto-capillary
balance leads to the formation of a thin
filament. The rest of the evolution
consists of an exponential thinning of the filament, best seen
in the logarithmic plot of Fig.\ref{fig2}. The slope drawn into
the figure corresponds to the theoretical prediction of section 4.

Structures very similar to Figure \ref{fig1} are shown in the
experiment of Figure \ref{clasen2new}
and have also been observed  by \cite{BVER81, CW01}
 for the decay of a liquid jet of polymer solution ejected from a nozzle. Good agreement with
numerical simulations of one-dimensional models very similar to ours have been reported
by \cite{Y93}. However, it is very difficult in practice to produce 
liquid cylinders without stretching the polymers, since there is 
considerable shear inside the capillary tube and the nozzle. 
On the other hand, the shear flow inside the capillary is very 
difficult to model in the 
framework of the present one-dimensional description. Furthermore,
each bead and ligament shown in Fig.\ref{clasen2new} corresponds to a `snapshot'
at a different elapsed convective time $\Delta \tilde{t} = L_{period}/v_{jet}$
Therefore, to compare quantitatively to experiments, we prefer to use a setup that allows for a
more quantitative description of the stretching history of the fluid column.

\subsection{Liquid bridge}

\begin{figure}
\begin{center} 
\psfig{figure=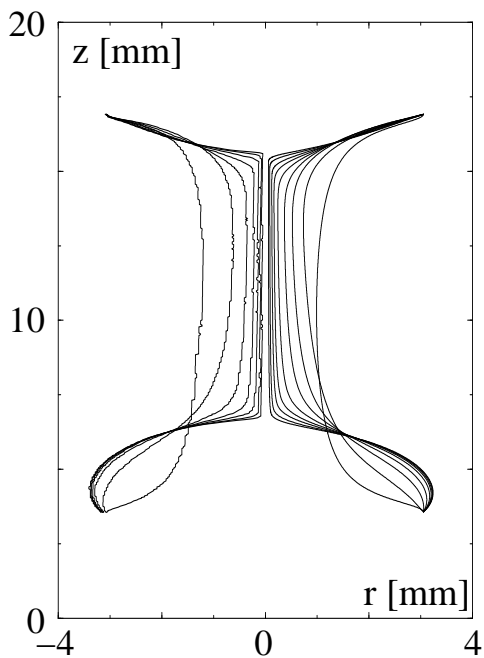,height=18cm,angle=-90} 
\end{center}
\caption{
A comparison between experimental profiles (left), obtained from
digitising the images of Fig. \ref{clasen1} and corresponding 
simulations (right), including the initial stretching and gravity (acting
downward).
The time difference between two consecutive images is 10 sec. The reduced 
variables corresponding to the experimental parameters are 
$\nu_s=193.2$, $\nu_p=29.04$, and $De = 296$. Note that there are 
{\it more} profiles from simulation to reach approximately the
same thread radius, corresponding to a slight overestimation of the 
experimental time scale.
}
\label{compare}
\end{figure}

\begin{figure}
\begin{center}
\psfig{figure=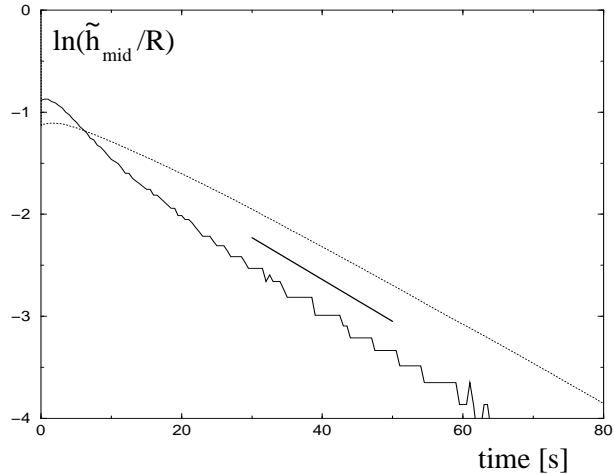,height=14cm,angle=-90}
\end{center}  
\caption{
The logarithm of the normalised minimum radius corresponding
to the experimental and theoretical profiles of Fig.\ref{compare}.
The thick line is the theoretical prediction for the slope 
$-1/(3\lambda)$.
  }
\label{midradius}
\end{figure}

We now turn to the capillary thinning and breakup 
 of a liquid bridge that has been subject
to a very rapid initial stretching. During this extensional step strain 
process, which lasts 0.05 sec, hardly any polymer relaxation takes place, so the 
exact manner in which the the plates are pulled apart is not 
very important. In fact, the initial stretching is well described
by a simple model neglecting any spatial structure (\cite{AM01}).
However, the simulation within the one-dimensional model is 
somewhat subtle, owing to a difficulty in imposing the boundary 
conditions at the end-plate. Namely, the no-slip boundary condition 
enforces a vanishing tangential velocity at the end-plate, and 
consequently $\partial v/\partial z = 0$, while this is not true 
in the one-dimensional model. Since the stretch rate 
$\partial v/\partial z$ is in fact large over most of the bridge,
this creates a thin boundary layer of the full three-dimensional 
flow near the end-plates. Failure to correctly implement this boundary 
layer leads to a detachment of the interface from the ends within
the lubrication model. 
Following \cite{STS00}, we have avoided this problem by introducing
a supplementary viscosity which strongly increases near the ends. 
This position-dependent viscosity has been constructed by matching
to a three-dimensional squeeze flow near a solid wall. The effective
``freezing'' of the fluid prevents any lateral slip along the bounding
wall.

Figures \ref{compare} and \ref{midradius} allow for a direct 
comparison between simulation and experiment. The first digitised
profile is taken just after cessation of stretching, after that 
a profile is shown every 10 sec. Theory and experiment show good
agreement in all the basic features of the flow, such as the sagging
under gravity and the formation of the thread. Two subtle differences 
can be seen: first, the timescale of the simulation is off by about
20\%, so 9 experimental profiles are shown but only 7 theoretical 
ones, at which point about the same minimum thread radius is reached.
This discrepancy, also seen in Figure \ref{midradius}, is quite
acceptable considering that no adjustable parameters were
introduced. Most of the difference stems from the early-time development,
while the asymptotic slope of both simulation and experiment agrees well 
with the theoretical prediction. 

The behaviour of $h_{min}$ at early
times is markedly different from that of a free jet, as discussed in
detail in \cite{AM01}. The rapid early decrease of $h_{min}$ 
seen in Figure \ref{fig2}
is absent, since stresses are already large owing to the initial stretching
of the liquid bridge. On the contrary, some of this initial stress 
has to relax before further thinning can start, as seen in the plateau
for the earliest times. Evidently there are some subtle features
of the experimental relaxation processes which are not modelled correctly
by our single-mode Oldroyd-B model. The initial stretch is also 
responsible for the absence of drops (``beads'') in the middle of 
the thread that formed on the free jet, cf. Figure \ref{fig1}. 
The reason is that 
the initial stretch is uniform, and this uniformity is conserved by 
the exponential stretching regime.

The second difference between the experimental and theoretical 
profiles of Figure \ref{compare}
is that at the same minimum thread radius the corner between the 
thread and drops at the end is {\it sharper}. We will return to 
this when we discuss the structure of the corner region in detail 
in section 4.2.

%% file: localapprox2.tex
\section{Early time asymptotics}

\subsection{Local description}
In this section we show that for early times $t \ll 1 \ll De$ 
the non-Newtonian
contribution to the stress behaves like that of a neo-Hookean elastic solid,
and can be expressed directly through the profile shape $h(z)$. This
proves the phenomenological argument of \cite{EY84}, in addition to
identifying the correct value of the elastic modulus of the material:
\begin{equation}
\label{elastic}
G = \nu_p/De.
\end{equation}
With the present scaling this is the same as the {\it elastocapillary}
number introduced in \cite{AM01} and can also be written $G =Oh(1-S)/De$.
Following the procedure in \cite{F02} we rewrite 
(\ref{b3}),(\ref{b4}) in Lagrangian
coordinates. This idea, introduced in \cite{R94}, is useful
since the stretch of a fluid element can be written as  $z_{\alpha} = 1/h^2$
(\cite {R94,E97}). This Lagrangian stretch $u\equiv z_{\alpha}$ is 
the derivative
of the particle position $z$ with respect to the  label $\alpha$ of the
particle in a reference configuration. Using the Lagrangian definition of
the fluid velocity $z_t=v$ one finds that $v_z=u_t/u$ and thus
\begin{eqnarray}
\sigma _{zz}+De\left( \frac{\partial \sigma _{zz}}{\partial t}-2\frac{u_{t}}{u%
}\sigma _{zz}\right) &=&2\nu _{p}\frac{u_{t}}{u}\;,  \label{b21} \\
\sigma _{rr}+De\left( \frac{\partial \sigma _{rr}}{\partial t}+\frac{u_{t}}{u}%
\sigma _{rr}\right) &=&-\nu _{p}\frac{u_{t}}{u}\;,  \label{b31}
\end{eqnarray}
which can be rewritten as
\begin{eqnarray*}
e^{-\frac{t}{De}}\frac{\partial }{\partial t}\left( \frac{1}{u^{2}}e^{\frac{t%
}{De}}\sigma _{zz}\right) &=&2G \frac{u_{t}}{u^{3}}\;, \\
e^{-\frac{t}{De}}\frac{\partial }{\partial t}\left( ue^{\frac{t}{De}}\sigma
_{rr}\right) &=&-G u_{t}\;.
\end{eqnarray*}

If we assume that the initial stresses are zero (otherwise they remain as
initial conditions) we can integrate the equations to find
\begin{eqnarray*}
\sigma _{zz} &=&2G u^{2}e^{-\frac{t}{De}}\int_{0}^{\tau}e^{\frac{\tau }{De%
}}\frac{u_{t}}{u^{3}}d\tau =-G u^{2}e^{-\frac{t}{De}}\int_{0}^{t}e^{%
\frac{\tau }{De}}\left( \frac{1}{u^{2}}\right) _{\tau}d\tau \;, \\
\sigma _{rr} &=&-G u^{-1}e^{-\frac{t}{De}}\int_{0}^{t}e^{\frac{\tau }{%
De}}u_{\tau}d\tau 
\end{eqnarray*}

Integration by parts yields
\begin{eqnarray}
\sigma _{zz} &=&-G +G \frac{u^{2}}{u_{0}^{2}}e^{-\frac{t}{De}%
}+\frac{G}{De} u^{2}e^{-\frac{t}{De}}\int_{0}^{t}e^{\frac{\tau }{De}}%
\frac{1}{u^{2}}d\tau \;, \label{szlagr} \\
\sigma _{rr} &=&-G + G \frac{u^{-1}}{u_{0}^{-1}}e^{-\frac{t}{De%
}}+\frac{G}{De}u^{-1}e^{-\frac{t}{De}}\int_{0}^{t}e^{\frac{\tau }{De}%
}ud\tau \;, \label{srlagr}
\end{eqnarray}
where the integrals are taken along Lagrangian paths, i.e. for constant
particle label $\alpha$. However, for $t \ll De$ the integral can be
neglected relative to $De$ and the exponential is close to unity.
For the case that the initial deformation of the bridge is small,
i.e. that $u_0$ is almost constant, (\ref{szlagr},\ref{srlagr})
only contains Lagrangian labels at the same time $t$, which can simply
be replaced by ordinary Eulerian coordinates by applying the inverse
transformation. Remembering also that with the scalings used throughout
this paper $u_0 = R_0 = 1$, we find
\begin{equation}
h^{2}\left( \sigma _{zz}-\sigma _{rr}\right) = G \left(
1/h^{2}-h^{4}\right) + O(t/De).
\end{equation}
This expression is the well-known result for the neo-Hookean rubber material
if we recognize that the Lagrangian stretch is $u = 1/h^2$.
Thus finally the equation for the velocity is
\begin{equation}  \label{early}
\;\frac{\partial v}{\partial t}+v\frac{\partial v}{\partial z}=\frac{1}{h^{2}%
}\frac{\partial }{\partial z}\left(h^{2}K +3\nu _{s}h^{2}\frac{\partial v}{%
\partial z} + G\left( 1/h^{2}-h^{4}\right) \right)
\;,
\end{equation}
where K is defined in (\ref{ber1}). Of course, (\ref{early}) can also
be confirmed directly by applying the limit $t \ll 1 \ll De$ to the Eulerian
equations (\ref{b3},\ref{b4}), but it is hard to guess a priori
without the use of Lagrangian coordinates.

\begin{figure}
\hspace{2cm}\psfig{figure=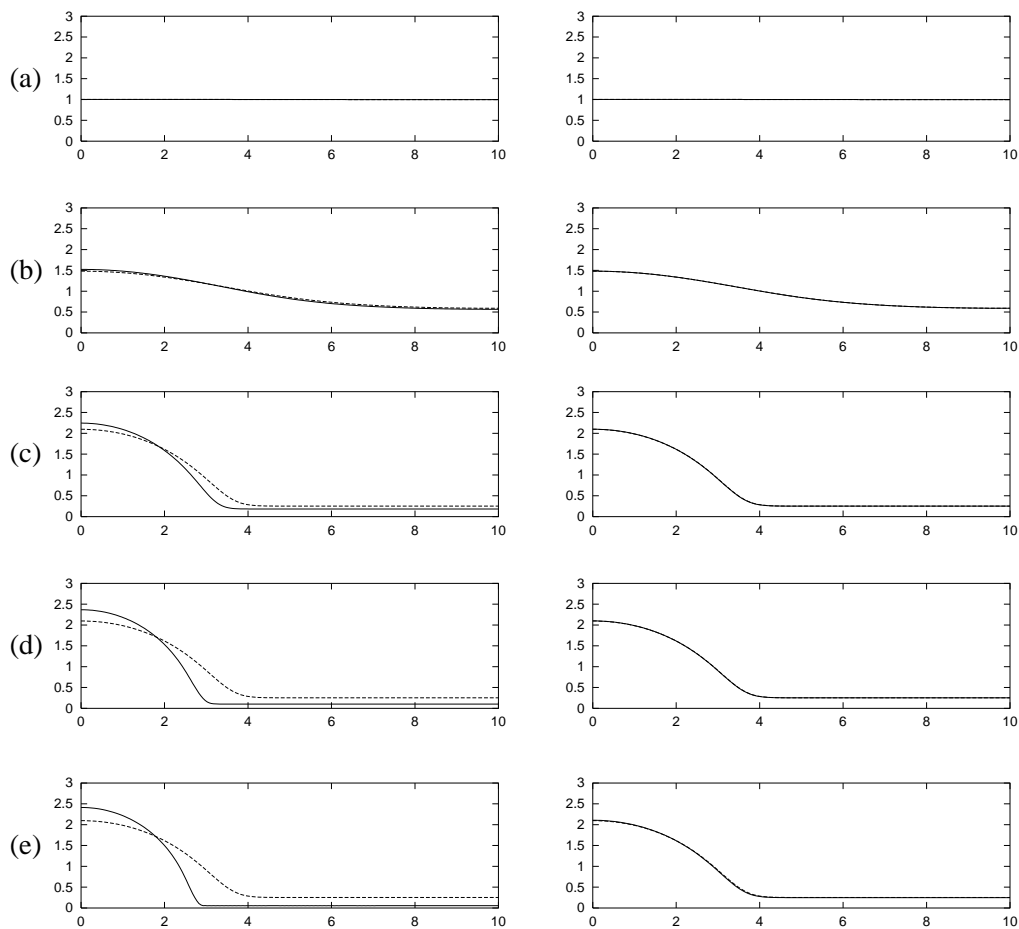}
\caption{
Comparison of interface profiles with period $L=20$ and
$\nu_s=0.79$ between  the
full equations (full) and the local approximation (dashed lines):
left column  $De = 94.9$ and $\nu_p=2.37$, and right column
$De = 9490$ and $\nu_p=237$.  The relative times for each profile are
(a) 31.6, (b) 79.1, (c) 158.1, (d) 316.2 and (e) 474.3.
Evidently, for $t \ll De$ the agreement is excellent (right column).
    }
\label{fig3}
\end{figure}

As a numerical test of the quality of the local approximation,
we performed two simulations similar to that of Fig. \ref{fig1},
but for two different {\it Deborah} numbers.
% XX Jens: this following caption doesnt match with the printed text 
%file I have? Is this
%    latest or is there a later one???
Fig. \ref{fig3} shows the evolution of the interface profile at different
times for $De = 94.9$, $\nu_p=2.37$ (left column),
and $De = 9490$, $\nu_p=237$ (right column).  The full
lines are the solution of equations (\ref{b1}-\ref{b4}), while the dashed
ones were obtained by replacing (\ref{b2}) by (\ref{early}).
For the simulation corresponding to  moderate Deborah number, the 
duration $t$ is of same order
of the relaxation time  $De$. And as we expected, the solution of the 
full equations and the one of
the local approximation  differ for times $t > De$.
 On the other hand, for the case of large Deborah number,  $t \ll De$, the
agreement of the two solutions is excellent.  Indeed,
we can hardly distinguish the  dashed lines from the full ones in the 
right column with our naked eyes.

Throughout the process of filament formation, we find almost
perfect agreement with the local model provided $De \gg 1$. Thus the
local approximation is an extremely useful tool to investigate the
early-time dynamics $t < 1 < De$ and the formation of the basic
beads-on-string structure. We will therefore study the stationary
filament solutions of the local model in the following subsection.
After the filament has formed, the neo-Hookean
elastic response of the local model leads
to a stationary profile, while the effects of fluid viscoelasticity
captured by the full equations leads to the filament continuing
to thin exponentially in time.

At low viscosities, $Oh < 1$, the initial evolution can be considerably
more complicated, but is still fully described by the local equations if
$De$ is sufficiently large. This is due to inertial effects also 
being important,
so as the thread is formed, fluid may rebound from the drops, and an
additional `secondary' drop forms in the middle of the filament.
This is also seen in experimental observations of jet breakup
(\cite{CW01}) and drop pinch-off (\cite{Coop02}). In addition, 
capillary waves may develop, which we describe in subsection 
\ref{Linear stability}.

\subsection{Static solutions}\label{sta}
In the neo-Hookean limit considered here, $De \rightarrow \infty$,
the elastic stresses never relax, so
at long times $t \gg De \gg 1$, surface tension is balanced by permanent
elastic stresses to form a stationary solution, which obeys the equation
\begin{equation}  \label{stat}
h^2 K + G \left(1/h^{2}-h^{4}\right) = C,
\end{equation}
which comes from integrating (\ref{early}) once while dropping
inertial terms.
Apart from the appropriate boundary conditions on $h$, (\ref{stat}) has to
be solved with the constraint of volume conservation
\begin{equation}  \label{vol}
\pi \int_0^L h^2 = V,
\end{equation}
which is needed to find all the constants of integration as well as $C$.

In particular, (\ref{stat}) allows for solutions with constant $h$
\begin{equation}  \label{cstat}
h + G\left(1/h^{2}-h^{4}\right) = C,
\end{equation}
where we have assumed a constant initial height $h_0=1$. The
solutions of equation (\ref{cstat}) correspond to the thin 
cylindrical threads of
constant radius described earlier. To determine their radius, one needs the
value of the constant $C$, which only follows from a solution of the full
system (\ref{stat}) with proper constraint (\ref{vol}). \cite
{EY84} have described a procedure which matches solutions of (\ref{cstat})
to almost circular drops in the limit of small elastic modulus $G$, 
from which the
thread radius can be determined analytically. Below we will reproduce
their calculation, which will turn out be relevant for the calculation
of thread thinning as well.

First, in the drop region the radius $h$ is of order one and one can 
neglect the terms
multiplied by $G$ in the limit we are interested in. Thus one is
left with the contribution from surface tension alone, and the solution is a
spherical drop,
\begin{equation}  \label{circle}
h(z) = R\sqrt{1 - (1+z/R)^2}.
\end{equation}
Here $R$ is the radius of the drop, which is set by the volume constraint (%
\ref{vol}), and must be a quantity of order one. The other constant of
integration was used to place the edge of the drop, which merges onto the
thread, at the origin. In the following we will assume that the filament is
to the right of the drop, in the region of positive $z$, while the matching
takes place in the neighbourhood of the origin.

To match the drop (\ref{circle}) onto a constant solution, we have to
analyse (\ref{stat}) in the limit $G \rightarrow 0$, guessing that in
the crossover region $h$ is still small. Thus $h^4$ can be neglected
relative to $1/h^2$ and we can eliminate the elastic modulus from the 
equation with
the scaling
\begin{equation}  \label{scaleps}
h=G^{1/3}H, \quad z=G^{1/3}Z, \quad C=G^{1/3}C_1,
\end{equation}
where now all rescaled quantities are of order unity and $H$ obeys the
equation
\begin{equation}  \label{statscal}
H^2K\{H\} + 1/H^{2} = C_1 .
\end{equation}
Denoting derivatives with respect to $X$ by primes and multiplying
(\ref{statscal}) by $H^{\prime}/H^3$ one finds the following
first integral,
\begin{equation}  \label{fint}
H^{\prime}=- \left[\frac{4H^2}{(C_1+2H^2/R_1-1/(2H^2))^2} - 1\right]^{1/2},
\end{equation}
where $R_1$ is a constant of integration. Here we have used
\[
\left (1 \over H^{2} \sqrt{1 + H_{z}^2}\right )^{'}
= - {H_{z} \over H \sqrt{1 + H_{z}^2}}
- {H_{z}H_{zz} \over H\sqrt{(1 + H_{z}^2)^3}} .
\]
The the constant $C_1$ and $R_1$ in the equation~(\ref{fint}) have to be
chosen so that the solution matches both the drop and the thread.

We begin with the matching to the drop for $z < 0$. For large $R_1$ there
is region with
\begin{equation}  \label{match1}
R_1^{1/2} \ll H \ll R_1
\end{equation}
in which (\ref{fint}) is approximated by
\[
H^{\prime}\approx -R_1/H
\]
with solution $H=\sqrt{2R_1|z|}$. By comparing with the expansion of the
drop profile (\ref{circle}) for small $|z|$ we find
\begin{equation}  \label{R1}
R_1 = R/G^{1/3},
\end{equation}
so $R_1$ is indeed large for small $\bar{\nu}$. In terms of the variable $z$,
the matching region (\ref{match1}) is defined by
\begin{equation}  \label{match2}
1 \ll |z| \ll R_1.
\end{equation}
On the other (thread) side of the problem ($z > 0$) the slope has to go to
zero and thus
\[
C_1 + 2H_0^2/R_1^2 - 1/(2H_0^2) = 2H_0,
\]
giving $C_1$ in terms of $H_0$. To finally find $H_0$, one has to make sure
that small perturbations around the constant solution do not grow. Putting
$H=H_0(1 + \delta)$, we find
\[
\delta^{\prime}\sim \left[(1 - 2H_0/R_1-1/(2H_0^3))\delta + O(\delta^2)%
\right]^{1/2}.
\]
For $\delta$ not to grow the expression in parentheses must vanish, and
since $R_1$ becomes large in the limit $G \rightarrow 0$, we are
finally left with $H_0=1/2^{1/3}$ and $C_1=3/2^{1/3}$. Thus the central
result of our analysis is that the thickness of the thread becomes
\begin{equation}  \label{thickness}
h_{thread} = \left(\frac{G}{2}\right)^{1/3}.
\end{equation}
For the parameters of Fig. \ref{fig3} we find $h_{thread} = 0.232$
from (\ref{thickness}), to be compared with the observed
value of $h_{min}= 0.253$. The quality of the approximation
quickly improves as the scale separation between thread thickness
and drop size becomes even more complete.

\subsection{Linear stability}\label{Linear stability}
The analyses of the previous two subsections can be used to perform
a linear stability analysis of the stationary threads that form in
the limit of very large $De$. This will be a good approximation on time
scales much smaller than $t < 1 \ll De$, over which the thread maintains
an almost uniform axial profile. This is a relatively simple calculation
to do, since the base solution is constant in both space and time.
Thus we put
\begin{equation}  \label{h0}
h = h_0+\widetilde{h}, \quad v = \widetilde{v},
\end{equation}
where $h_0$ is constant and $\widetilde{h}$, $\widetilde{v}$ are small
perturbations around the base solution. Then the linearised equations turn
out to be:
\begin{eqnarray*}
&&\frac{\partial\widetilde{h}}{\partial t}+\frac{1}{2}h_0\frac{\partial
\widetilde{v}}{\partial z} =0 \\
&&\frac{\partial \widetilde{v}}{\partial t} =\frac{1}{h_0^{2}} \frac{%
\partial }{\partial z}\left( h_0^{2}\widetilde{h}_{zz}+3\nu _{s}h_0^{2}\frac{%
\partial \widetilde{v}}{\partial z}+\left( 1-\frac{2}{h_0^{3}}G
-4h_0^{3}G \right) \widetilde{h}\right) ,
\end{eqnarray*}
which can be combined to the following equation for $\widetilde{h}$ alone:
\begin{equation}
\frac{\partial ^{2}\widetilde{h}}{\partial t^{2}}=-\frac{1}{2}h_0\frac{%
\partial ^{4}\widetilde{h}}{\partial z^{4}}+3h_0\nu _{s}\frac{\partial^{3}%
\widetilde{h}}{\partial z^2\partial t}-\frac{1}{2h_0}\left( 1-\frac{2}{%
h_{0}^{3}}G -4h_0^{3}G \right) \frac{\partial ^{2}\widetilde{%
h}}{\partial z^{2}}.
\end{equation}

Using a wave ansatz of the form $\widetilde{h}(z,t)=\epsilon \exp{%
i(kz-\omega t)}$ we find the dispersion relation
\begin{equation}  \label{dispersion}
\omega ^{2}=-\frac{k^2}{2h_0}(1 - h_0^2k^2) + \frac{k^2 G}{h_0^4}(1 +
2h_0^6) + 3i\nu_s h_0\omega k^2.
\end{equation}
The first term on the right comes from surface tension and describes the
usual Rayleigh-Plateau instability of a Newtonian liquid jet. If we
disregard the polymer contribution we find complex $\omega$ for $k h_0 < 1$ and
thus exponential growth of perturbations. The presence of elastic 
stresses shifts the
critical value of $k$ to slightly smaller values of $k$, thus stabilising
the flow. If however this critical value as well as the maximum growth rate
are compared to a Newtonian fluid with the same zero shear rate 
viscosity $\eta_0$, one
finds that the instability grows {\it faster} 
(e.g. \cite{Mid65, GYPS69,CDK99}).

In the thin thread region described by (\ref{thickness}), on the other hand,
one is effectively dealing with an elastic medium and can see the
propagation of elastic waves. Only taking into account the elastic
contributions one finds
\begin{equation}  \label{wave}
\omega = \pm\left[\frac{\bar{\nu}k^2}{h_0^4} - (3\nu_sh_0k^2/2)^2\right]
^{1/2} + 3i\nu_sh_0k/2.
\end{equation}
Hence for small viscosities there is little damping and waves are very
easily excited on the threads, which causes considerable numerical problems
to properly resolve them. In the limit of zero viscosity the elastic
wave speed is given by
\begin{equation}  \label{speed}
v_{el} = \frac{\sqrt{G}}{h_0^2}.
\end{equation}

%% file: expon2.tex
\section{Late time asymptotics}

\subsection{Thread thinning}
The formation of threads described in the previous section is a result of
the interplay of surface tension and elastic forces. Any additional thinning
of the thread results in an extensional flow, resisted by further stretching
of the polymers. On times longer than $De$
however the stress gradually relaxes, and the thread thins at an
exponential rate $\beta$. This rate is easily determined from a balance of
surface tension and elastic forces in (\ref{b2}), assuming a spatially
constant profile (e.g. \cite{BVER81,R95,EH97})

\begin{equation}  \label{thin}
h(z,t) = h_0\exp(-\beta t).
\end{equation}

 From volume conservation (\ref{b1}) one finds that the extension rate $%
\partial_z v = 2\beta$ in the thread is constant. The exponential growth of
the axial stress $\sigma_{zz}$ is described by (\ref{b3}), and assuming a
spatially constant $\sigma_{zz}$ one immediately finds

\begin{equation}  \label{sthin}
\sigma_{zz}(z,t) = \sigma_0\exp[(4\beta-De) t].
\end{equation}

The radial stress $\sigma_{rr}$ {\it decreases} exponentially and does not
figure in the balance. Remembering that capillary pressure is balanced
with $\sigma_{zz}$ in (\ref{b2}) and rewriting the pressure gradient
according to (\ref{ber1}) one finds

\begin{equation}  \label{balance}
h + h^2\sigma_{zz} = C,
\end{equation}

performing one spatial integration. For the balance (\ref{balance}) to be
consistent $\sigma_{zz}$ must grow like $1/h$, and thus $\beta=1/(3De)$,
implying that the constant of integration $C$ itself decays like
\begin{equation}
C = a_1\exp[-t/3De].
\end{equation}
This means that the thinning rate of the thread is directly related to the
time scale of the polymer, providing for a convenient experimental probe.

\begin{figure}
\hspace{3cm}\psfig{figure=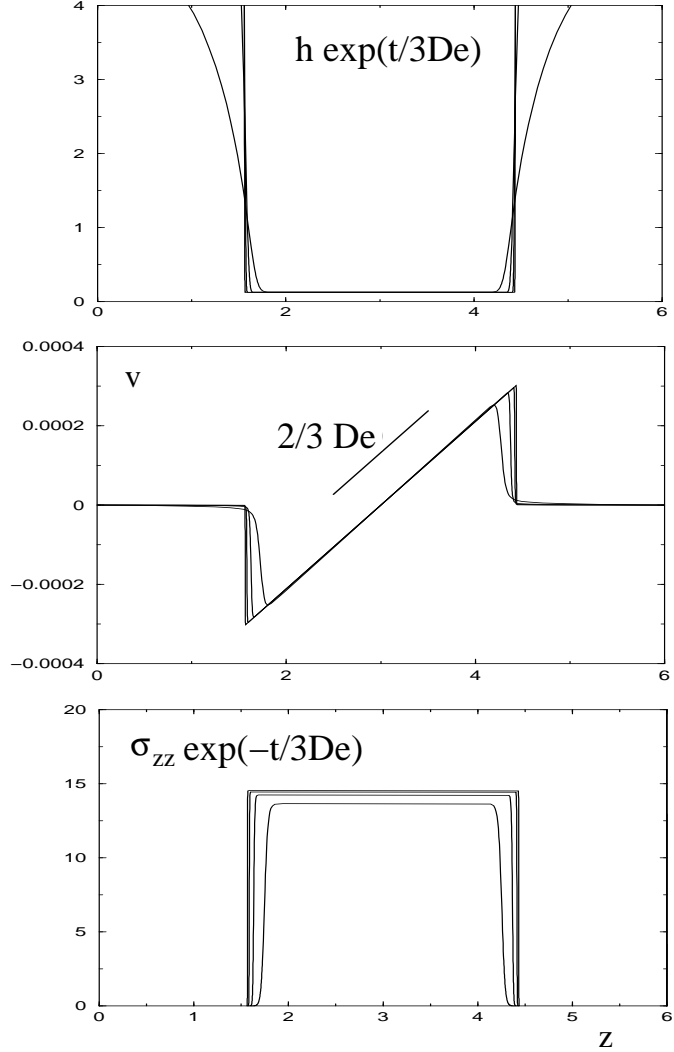,height=14cm}
\caption{
Profiles of the radius, the velocity, and the axial stress, normalised
by their values in the thread, in the regime of exponential thinning.
The parameters are $De = 3164$ and $\nu_s=\nu_p = 95.6$. The extension
rate inside the thread is $2/(3De)$.
}
\label{thread}
\end{figure}

To obtain a clearer physical picture, we plot in Figure \ref{thread} the
thread radius, normalised by the exponential factor $\ell=\exp[-t/3De]$, the
velocity in the thread, and $\ell\sigma_{zz}$ for a number of different
times during the thinning. As predicted, $h(z)$ and $\sigma(z)$ are very
nearly constant over the thread, and they collapse nicely as anticipated by
the above scaling laws. Furthermore, the extension rate $\partial_z v$ has
the constant positive value $2\beta$ inside the thread, expressing the fact
that fluid is expelled from
it. In response, the stress $\sigma_{zz}$ grows
to large {\it positive} values, so both contributions on the left-hand side
of (\ref{balance}) are always positive. This means that the constant of
integration in (\ref{balance}) (which below we will compute explicitly in
the limit of large $De$) has to be kept for a consistent balance and is
needed to determine the stress and thus the extensional viscosity from a
measurement of the thread radius. For large $De$ this
makes the stress twice as large (apart from a difference in sign) as
determined from the customary direct balance between capillary and elastic
stresses (e.g. \cite{BVER81,EH97,AM01}).

Since the total length $L$ of the thread is a constant quantity, the maximum
value of the velocity
\begin{equation}  \label{vmax}
v_{max} \approx \frac{L}{3De}
\end{equation}
behaves like the inverse of the Deborah number, and is thus small in the limit
that is the chief focus of this paper. A number of experiments (e.g. 
\cite {BELR,AM01})
have confirmed the prediction (\ref{thin}). In particular
in \cite{AM01} the relaxation time was determined independently,
and the thinning
rate was found to conform with the prediction $\beta=1/(3De)$. At both ends
of the thread, the velocity and the stress fall to zero very sharply, and
the profile merges onto a static drop with radius $R$. Next we will focus on
this transition region, whose scale is set by $\ell=\exp(-t/3De)$.

\subsection{The corner region}

According to the scalings found in the previous subsection, which implies
the existence of a small length scale $\ell=\exp(-t/3De)$, it is natural to
look for solutions of (\ref{b1}-\ref{b2}) of the form

\begin{eqnarray}
&& h(z,t) = \ell\overline{h}(\overline{z},t)  \nonumber \\
&& v(z,t) = \overline{v}(\overline{z},t)  \label{line} \\
&& \sigma _{zz}(z,t) =\ell^{-1}
\overline{\sigma}_{zz}(\overline{z},t) ,  \nonumber
\end{eqnarray}

where $\overline{z} = \ell^{-1} (z - z_0)$. The origin $z_0$ must
asymptotically lie in the similarity region. A convenient choice
is the position of the extremum of the velocity in the limit
$\ell\rightarrow 0$. Since $\sigma_{rr}$ is exponentially
small inside the thread, it can be left out of our analysis.
Thus the equations for $\overline{h},
\overline{v}$, and $\overline{\sigma}_{zz}$ are

\begin{eqnarray}
&& 2\ell\overline{h}\dot{\overline{h}} - \frac{2\ell}{3De} \left(\overline{h}%
^{2} - \overline{z}\overline{h}\overline{h}^{\prime}\right) + \left[%
\overline{v}\overline{h}^2\right]^{\prime}= 0 \nonumber \\
&& \overline{h}^{2}\left(\ell^2\dot{\overline{v}} + \frac{\ell^2\overline{z}%
}{3De}\overline{v}^{\prime}+ \ell\overline{v}\overline{v}^{\prime}\right) = %
\left[\overline{h}^2 K\{\overline{h}\} + \overline{h}^{2} (3\nu_{s}
\overline{v}^{\prime}+ \overline{\sigma}_{zz})\right]^{\prime}  \label{fse}
\\
&& \ell\dot{\overline{\sigma}}_{zz}+\frac{\ell}{3De} (\overline{\sigma}_{zz}+%
\overline{z}\overline{\sigma}_{zz}^{\prime})+ \overline{v}^3\left[\frac{%
\overline{\sigma}_{zz}}{\overline{v}^2}\right]^{\prime}= \frac{\ell}{De}%
(2\nu_p\overline{v}^{\prime}-\overline{\sigma}_{zz}),  \nonumber
\end{eqnarray}

where the prime refers to differentiation with respect to the
similarity variable $\overline{z}$ and the overdot indicates
differentiation with respect to time. Towards the thread our scaling
ensures that $\overline{h},%
\overline{v}$, and $\overline{\sigma}_{zz}$ tend toward the constants $h_0,
v_0$, and $\sigma_0$, respectively, as $\overline{z}\rightarrow \infty$.
(Without loss of generality, we assume that the thread is to the right of
the transition region.) The length scale $\ell$ becomes exponentially small
for $t\rightarrow \infty$ and for any finite $\overline{z}$, terms
proportional to $\ell$ or $\ell^2$ may be dropped. Looking for
time-independent solutions of (\ref{fse}) and integrating once, we find
\begin{eqnarray}
&& \overline{v} \overline{h}^{2} = v_0h_0^2  \nonumber \\
&& \overline{h}^2 K\{\overline{h}\} + \overline{h}^{2} (3\nu_{s} \overline{v}%
^{\prime}+ \overline{\sigma}_{zz}) = h_0 + h_0^2\sigma_0  \label{ar6} \\
&& \overline{\sigma}_{zz}/\overline{v}^{2} = \sigma_0/v_0^2 .  \nonumber
\end{eqnarray}

\begin{figure}
\begin{center}
\psfig{figure=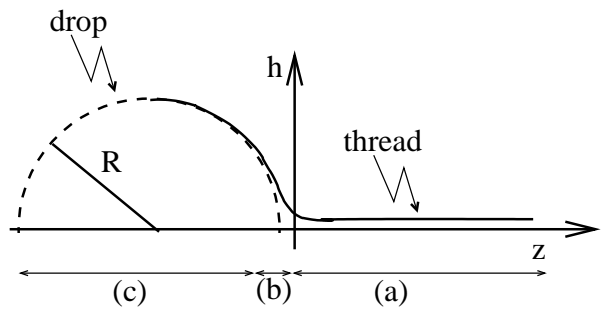,height=5cm}
\end{center}
\caption{
A schematic of the different regions near the corner.
}
\label{scheme}
\end{figure}

Eliminating $\overline{v}$ and $\overline{\sigma_{zz}}$, we end up with

\begin{equation}
\overline{h}^2 K\{\overline{h}\} + \overline{h}^{2}(-a_2\overline{h}%
^{\prime}/\overline{h}^3 + a_3/\overline{h}^4) = a_1,  \label{ecu}
\end{equation}

where $a_1 = h_0 + h_0^2\sigma_0$,$a_2=6\nu_s v_0h_0^2$, and $%
a_3=\sigma_0h_0^4$. In general, equation (\ref{ecu}) is no longer valid
inside the drop, because here the profile $h(z,t)$ is almost stationary,
thus $\overline{h}$ varies like $\ell^{-1}$, implying that the time
derivative $\dot{\overline{h}}$ is proportional to $1/(\ell De)$. In
addition, the similarity variable $\overline{z}$ is also of order $1/\ell$
inside the drop. Therefore we have to distinguish three different regions,
as shown in schematically in Figure \ref{scheme}.

\begin{itemize}
\item[(a)]  The similarity region near the thread described by (\ref{ecu}).
\item[(c)]  The solution inside the almost spherical drop,
where the velocity and stress fields relax towards zero.
\item[(b)]  A transition region connecting the dynamical region (a) with the
static region (c).
\end{itemize}

However in the limit of large $De$ which is the main focus of this paper,
velocities are of the order of $1/De$, and the contribution
$-a_2\overline{h}^{\prime}/\overline{h}$ in (\ref{ecu}), which
comes from the viscous stress, can be neglected. The same is true
for the left hand side of the second equation (\ref{fse}),
as we will show in more detail in the appendix.
Physically this limit corresponds to a situation where the fluid flow
becomes irrelevant to the problem, which is dominated by surface tension
and elastic contributions. Since the solution of (\ref{ecu}) matches
onto the drop, $\overline{h}$ must become as large as desired for
a sufficiently large $\overline{z}$. But this means that the contribution
from $\overline{h}^2\overline{\sigma}_{zz} = a_3/\overline{h}^2$
is negligible compared to the constant $a_1$ outside of the similarity
region. Thus although $\overline{\sigma}_{zz}$ can no longer be written
as a simple function of $\overline{h}$ in regions (b) and (c),
it does not contribute to leading order outside of the similarity region (a).
In summary, this implies that (\ref{ecu}) with $a_2=0$ is valid
{\it everywhere} in regions (a)-(c).

The constant $a_3$ can be scaled out by putting

\begin{equation}  \label{hbscal}
\overline{h}(\overline{z},t) = a_3^{1/3}g
(\zeta \equiv \overline{z}/a_3^{1/3},t),
\end{equation}

and so we finally find

\begin{equation}  \label{ecug}
g^2K\{g\} + 1/g^{2} = C_1
\end{equation}

where $C_1 = a_1/a_3^{1/3}$.

Equation (\ref{ecug}) is identical to (\ref{statscal}),
describing the stationary thread
in the local approximation. The matching described in section 3.2

can thus be applied with very slight  modifications. Multiplying (\ref{ecug})
by the integrating factor $g'/g^3$ and integrating once, we obtain

\begin{equation}
g^{\prime }=-\left[ \frac{4g^{2}}{(C_{1}+2g^{2}/R_{1}(t)-1/(2g^{2}))^{2}}-1%
\right] ^{1/2}.  \label{j14}
\end{equation}

We seek values of $C_{1}$ and $R_{1}(t)$ such that

\begin{eqnarray}
g(\zeta,t)\sim Const.\ \  {\rm when }\ \zeta \gg 1, \\
g(\zeta,t)\sim (2R)^{\frac{1}{2}}(-\ell ^{-1}a_{3}^{-1/3}\zeta )^{\frac{1}{2}}
\ \ {\rm when }\ \zeta \ll -1,
\end{eqnarray}

implying matching with the filament and the drop respectively.
In order to match to the
stationary drop of radius $R$ we put $R_1(t) = R / (\ell a_3^{1/3})$ in
analogy to (\ref{R1}), and in the limit of $t\rightarrow \infty$ we arrive at
the equation

\begin{equation}  \label{gequ}
g^{\prime}=-\left[\frac{4g^2}{(C_1 -1/(2g^2))^2} - 1\right]^{1/2}.
\end{equation}

The procedure to match with the filament described in section~3.2
establishes that $C_1 = 3/2^{1/3}$ and

\begin{equation}  \label{amp}
\sigma_0 = 2/h_0, \quad a_1 = 3h_0.
\end{equation}

The transition region (b) between the drop and the thread is by virtue of
(\ref{match2}) defined by

\begin{equation}  \label{match3}
1 \ll |\overline{z}| \ll R / \ell.
\end{equation}

In this region the solution is

\begin{equation}  \label{match4}
g=\sqrt{2R_1|\overline{z}|},
\end{equation}

which in physical variables reads $h=\sqrt{2R |z|}$ and thus matches to the
stationary drop. The readily measurable amplitude $h_0$ of the thread
thinning rate remains the only free parameter of the theory. Remarkably,
the amplitude of the stress is exactly {\it twice} the value expected
from a naive balance of capillary and elastic forces.

Integrating (\ref{gequ}) with initial radius $1/2^{1/3}$ given by the form of
the solution (\ref{gequ}) (plus an arbitrarily small perturbation) one finds
the universal profile shape $g$. The similarity profile $\overline{h}$ is then
given by (\ref{hbscal}) with $a_{3}=2h_{0}^{3}$. From Figure \ref{thread} one
reads off that $h_{0}=0.1275$ for our simulation with $De=3164\gg 1$, and the
numerically computed profiles in the corner region can easily be
converted to give the rescaled profile $g(\zeta)$.
\begin{figure}
\begin{center}
\psfig{figure=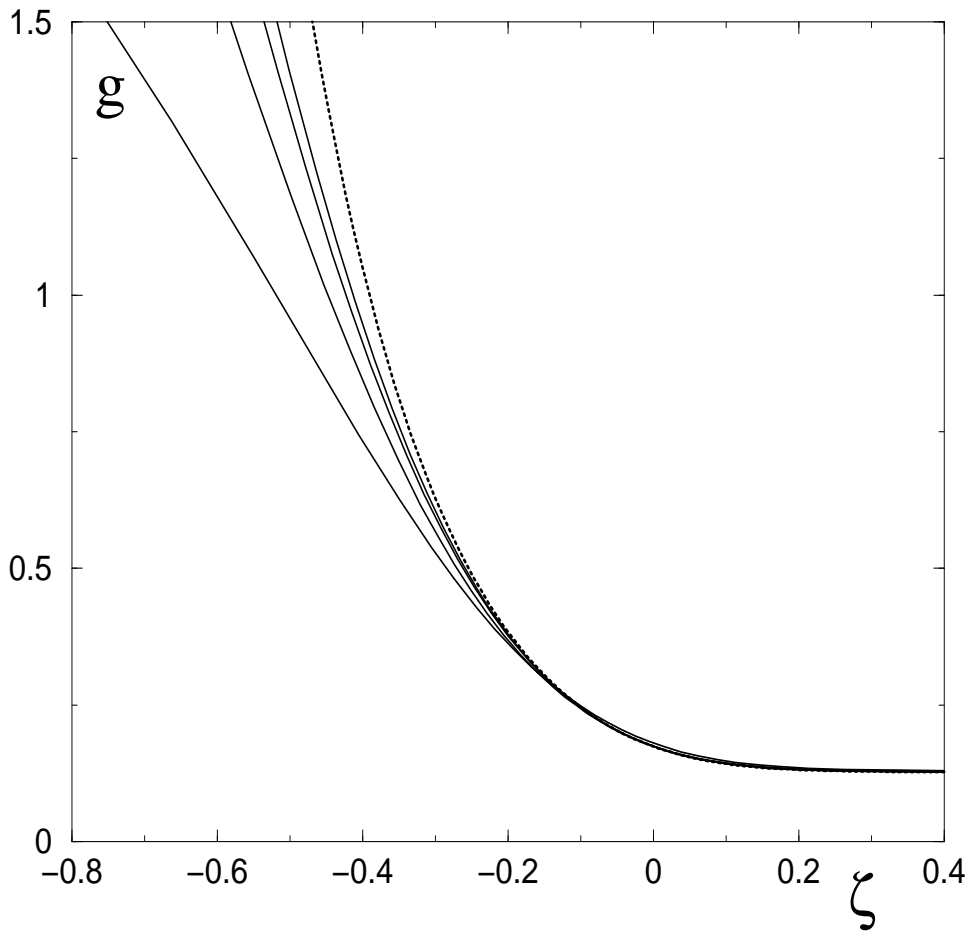}
\end{center}
\caption{
A comparison of our similarity theory for large $De$ (dashed line)
with rescaled profiles from a numerical simulation with parameters
as in Figure \ref{thread}, at $log_{10}(h_{min}) = -1.5,-2,-2.5,$ and
$-3$. The dashed line was obtained from integrating (\ref{gequ}),
and rescaling according to (\ref{hbscal}), with $\zeta=\bar{z}/a_3^{1/3}$.
}
\label{sim}
\end{figure}
The similarity profiles, computed from the numerical results of
Figure \ref{thread} for four different times (solid lines),
are compared with the theoretical prediction (
dotted line) in Figure \ref{sim}.
As $t$ grows, the rescaled profiles converge nicely to the
theoretical prediction.

Finally, we compare our theoretical results to experiments.
To that end, high resolution images of experimental profiles
were taken by focusing a video microscope on the corner region.

\begin{figure}
\begin{center}
\psfig{figure=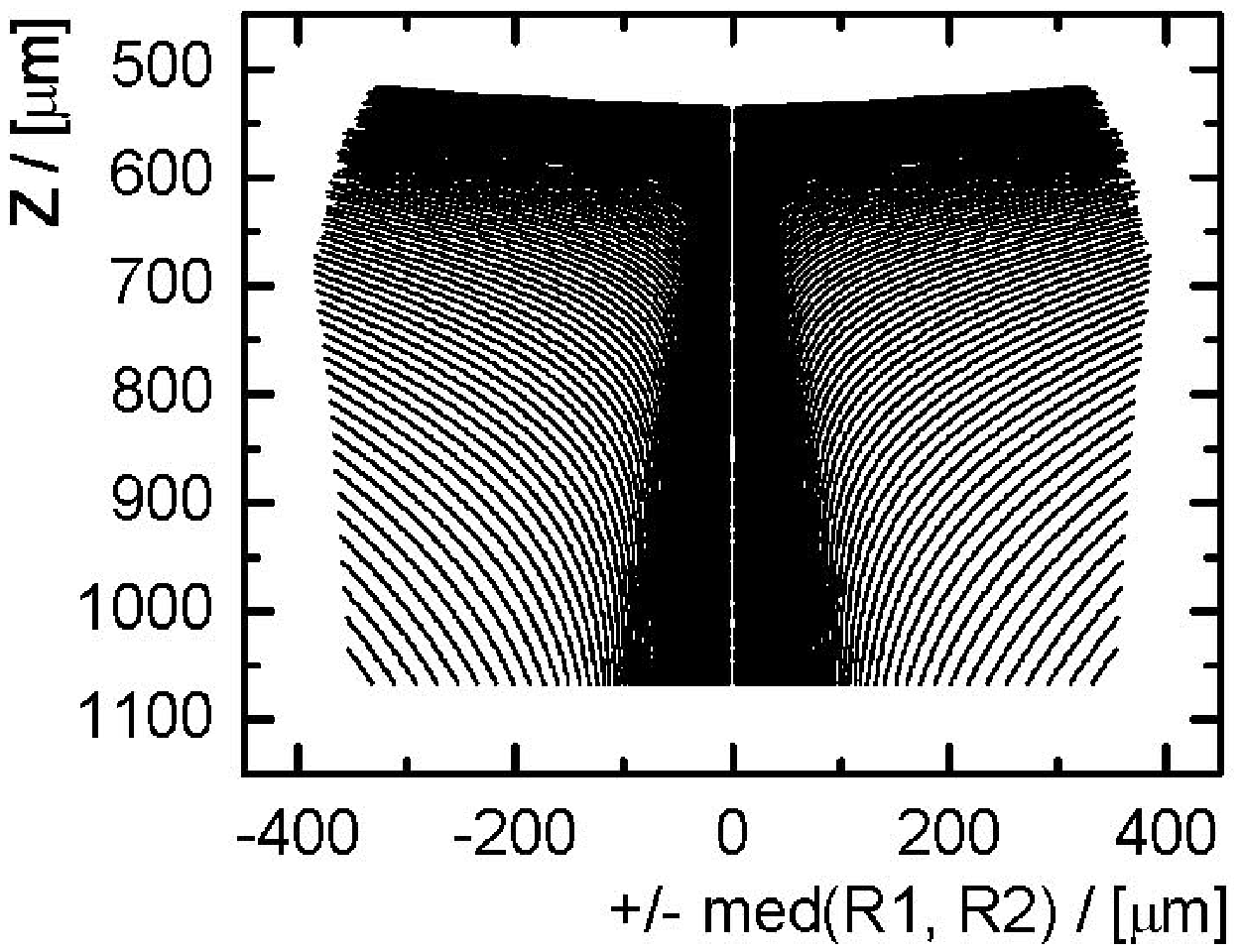}
\end{center}
\caption{
An experimental closeup of the upper corner of the liquid
bridge. Shown is the same experiment as Figure \ref{clasen1}, but under a
microscope. The time interval between two consecutive profiles is 0.5 sec.
   }
\label{expcorner}
\end{figure}

\begin{figure}
\begin{center}
\psfig{figure=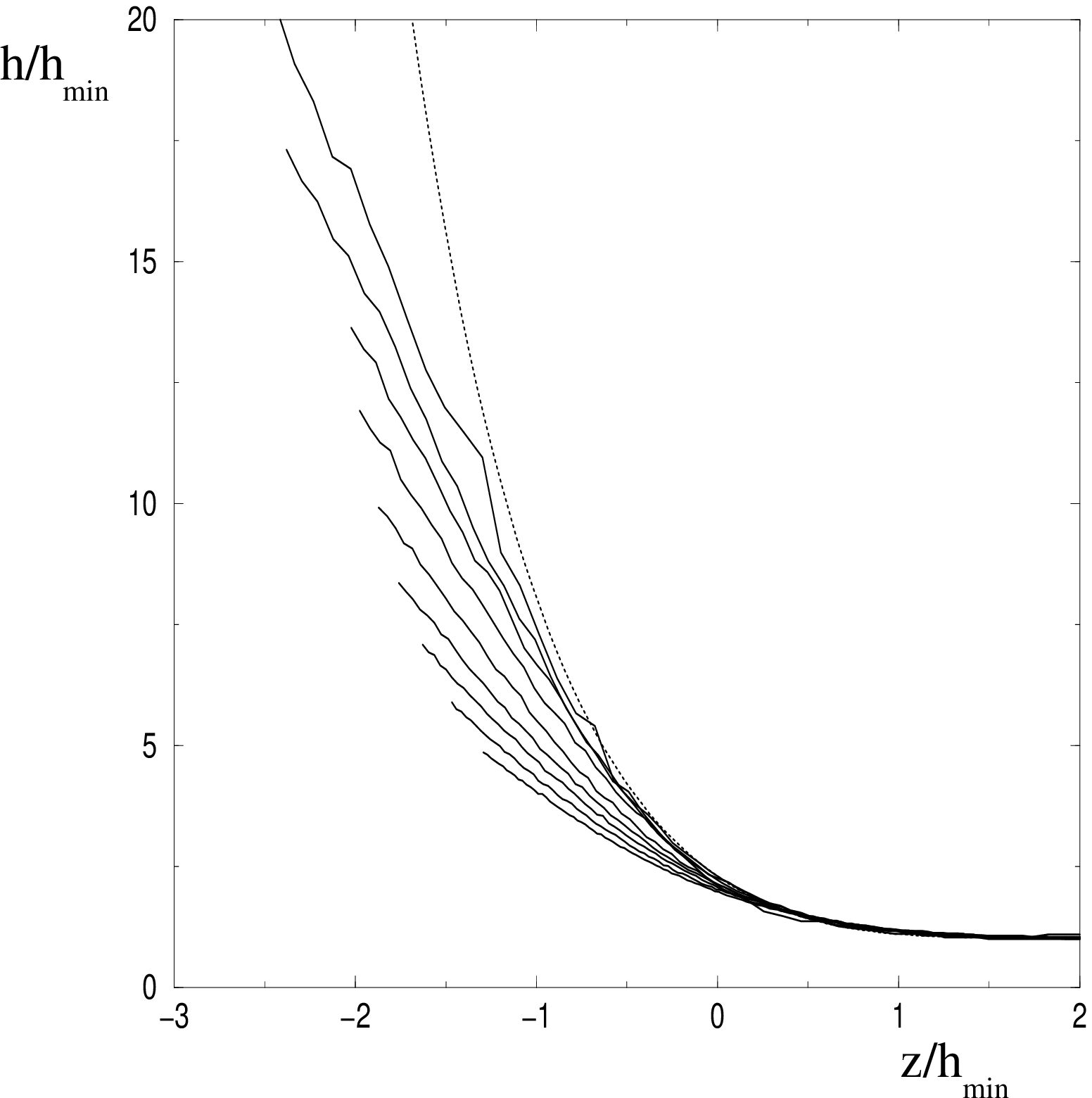,height=8cm}
\end{center}
\caption{
The last 9 profiles of Figure \ref{expcorner}, with both axes rescaled
by the minimum thread radius. The dotted line corresponds to
the asymptotic theory also shown in Figure \ref{sim}, but with the
horizontal axis {\it shrunk} by a factor of 0.5.
}
\label{ccorner}
\end{figure}

Figure \ref{expcorner} shows a sequence of digitized images
of the upper corner, which becomes increasingly sharp as the thread
thins. To test our similarity theory, the profiles
corresponding to the latest stages of pinching were rescaled using the
minimum thread radius
for both axes. As shown in Figure \ref{ccorner}, the experimental
profiles converge nicely onto a master curve, in very much the
same way as the computed profiles of Figure \ref{sim} do. However,
the experimental profiles turn out to be {\it sharper} than theory
predicts. Thus the asymptotic result given by theory needs to be
shrunk horizontally by a factor of 0.5 to fit the experimental curves.
The resulting curve is given as the dotted line in Figure \ref{ccorner}.

The constant $h_0$ is the only remaining adjustable parameter
for the description of the corner region. In the case of the
free jet, cf. section 2.2, it can in fact also
be estimated in the limit that the relaxation time $De$ of the polymer
is much larger than the time $t_{fil}$ needed to form the primary
filament. Namely, equating the thread thickness (\ref{thickness})
given by the local theory of section 3 with
$h_{min}=h_0\exp(-t_{fil}/3De)$ one finds

\[
h_0=\exp(t_{fil}/(3De))\left(\frac{G}{2}\right)^{1/3}.
\]

Thus in the limit that $De$ is much larger than $t_{fil}$
one simply has

\begin{equation}  \label{h0amp}
h_{min}(t) = \left(\frac{G}{2}\right)^{1/3}\exp(-t/3De),
\end{equation}

which is smaller by a factor of $2^{-1/3}$ than the result given in
\cite{BELR} and \cite{EH97}, based on a direct balance of elastic 
and surface tension forces. Fitting
a straight line to the exponential thinning regime of Figure \ref{fig2},
we obtain $h_0 = 0.247$, in excellent agreement with the theoretical
prediction of (\ref{h0amp}), $h_0 = (G/2)^{1/3} = 0.232$. 

%% file: conclusion2.tex
\section{Discussion}

In the present paper, confining ourselves to the simplest
possible model for a dilute polymer solution, we have reached a
rather complete description of the formation and subsequent
development of the beads-on-string structure. Some details,
of course, remain to be elaborated.

At low viscosities or high surface tension, $Oh \ll 1 \ll De$
our preliminary studies (e.g. \cite{FL03}) indicate
that a complex succession of beads may be generated owing to
inertial effects. Namely, fluid that rebounds from one of the
large drops leads to capillary waves that travel on the thread,
and can form one or more additional drops.
After a sequence of drops has formed, these drops are subject
to  possible merging, or draining processes.
%In \cite{FL03} it is conjectured that the final stationary state of
%the local model only allows for a {\it single} large drop, except
%for the non-generic case of several drops being exactly equal.

A similar phenomenon of secondary drop formation has been described
in \cite{CDK99} and is referred to as `recoil'. This recoil leads to the
formation of a secondary filament between the secondary and main
drops which, as claimed in \cite{CDK99}, exhibits
a series of subsequent instabilities. However, for the parameters
given in \cite{CDK99} we found no indication for the formation
of a secondary filament, but only for significantly lower viscosity
or higher surface tension. Even if we do find a secondary filament,
its presence is due to inertia, while \cite{CDK99} claim their
instability to be independent of inertial effects. Never did
we find any indication of another instability leading to a third
generation filament. These observations have been checked using both
our implicit and explicit codes, making sure that the resolution
was sufficient in the thin transition region between filament and
drop. Insufficient resolution could occasionally lead to instabilities
that vanished as resolution was increased.

Still within the framework of the Oldroyd-B model, but at more moderate
values of $De \sim O(1)$, viscous flow effects would have to be
taken into account (cf. Fig.\ref{clasen3new}). As indicated above, the
velocity in the corner region is proportional to $1/De$, thus in the
limit of large $De$, viscous  effects drop out and only capillary and
elastic forces remain.  In the presence of viscous forces, equation
(\ref{ecu}) can be applied for any finite value of the similarity
variable $\overline{z}$.  However, we have not yet succeeded in matching
the solution described by (\ref{ecu}) with the stationary drop in the
general case.  The reason is that whilst
in the similarity region close to the corner (between the thread and the
drop) everything can be described by the thread profile $\overline{h}$
alone, this is no longer possible in the transition region towards
the stationary drop. Instead, the coupled system involving
the drop shape, the velocity field, and the stress has to be
treated. Indeed, even in the limit of large $De$ where matching can be
achieved using $h$ alone, analytical computation of $v$ and $\sigma_{zz}$
is far from trivial (\cite{Fun}).

A result of the present similarity description is that
for very long times the drop develops into a perfect sphere, while
the thread radius shrinks to zero. This means the profile at the
point where the thread meets the drop develops into a right angle.
The same conclusion, but using a completely different approach
without making use of the self-similar structure of the equations,
was reached in \cite{R94}. The series of experimental profiles
shown in Fig.\ref{expcorner} supports this conclusion.

Unfortunately, the fact that the slope of the profile becomes
increasingly large in the corner means that the lubrication equations,
used throughout this paper, are no longer valid. While it does
not seem likely that this fact changes the scaling found in the present
paper, it is clear that the precise shape of the profile will be
somewhat different for the full equations. Full numerical simulations 
will also be quite challenging to perform with the accuracy required 
to resolve the corner region in the presence of strong elastic effects. 
However, one could use
the self-similar scaling proposed in the present paper to reduce the 
dynamical 3-D equations to a time-independent equation, analogous to 
(\ref{ar6}). In rescaled coordinates, the scale of the corner 
is of order one, and a solution of the problem could be attempted 
numerically. For the
near future, the most promising prospect appears to be to compare with
experiments with very long, well characterised polymer solutions.
Note that even within a full 3-D theory, the correct minimum radius 
cannot be obtained from a direct balance of elastic and surface 
tension forces, as it was done in earlier works (\cite{EH97,BELR}).
The reason is that both contributions have the same sign (compare
(\ref{balance})), and thus cannot possibly balance. Instead, a 
matching to the drop has to be performed in order to determine 
the total tension in the thread. 

Finally, there are myriad effects associated with departures from
the Oldroyd-B model, some of which have already been incorporated into
the description of thread thinning. For example, even a perfectly
mono-disperse polymer solution is expected to be described by a
{\it spectrum} of time scales, inherent in the complex relaxation
dynamics of even a single polymer strand (\cite{pgg79}).  This is not
expected to be a major limitation for the present theory, however.
Bead-spring models of dilute polymer solutions indicate that there is a
simple recursion formula  relating the longest relaxation time $\lambda$
and higher order relaxation processes, of the form $\lambda_k =
\lambda/k^m$, with $m = 1.5,2$ for the Zimm or Rouse models respectively
(\cite{BAH87}).
The local rate of stretching in the cylindrical thread arising
from the elastocapillary balance is sufficiently weak
(in dimensional form $\dot{\varepsilon} =2/(3\lambda)$), that
all other modal contributions to the stress (of the form of eqs. 2.3, 2.4)
  will decay away and not contribute to the elastic stress
since the convective terms will be of the form $(2De) v'/k^m < 1$.
A detailed  analysis of the effect of such a spectrum of relaxation times
was carried out in \cite{EH97}, but using an incorrect form of
the force balance, eq. (\ref{balance}).

   The effect of the finite extensibility of a polymer chain
is expected to change the picture in Fig.\ref{clasen3new}, however,
since it bounds the maximum elastic stress that can be
exerted by the polymer chain.  This is
modelled  theoretically by the presence of nonlinear terms in the
constitutive  equation for the polymeric stresses, which limit their
growth. Thus while  the exponential thinning described in this paper would
formally  lead to breakup only in infinite time, a real polymer thread
does in  fact break. Experimental observations of the departure from the
exponential law are found for example in \cite{BELR} and
\cite{AM01}. The theory in \cite{BELR}
differs from the conventional one in that it credits the  {\it
degradation} of polymers for the departure from the exponential law.

%Unfortunately, there is no accepted nonlinear law describing
%finite extensibility, but an overview is given in \cite{BAH87}.
%In a recent paper, \cite{R01} finds a similarity solution for
%finite-time  breakup in the framework of the so-called Giesekus model,
%which  amounts to assuming a restoring force that is quadratic in the
%polymer extension. The Giesekus model is subject to criticism,
%because it allows for polymer extensions which are proportional
%to the extension rate, while the maximum extension is rather a
%geometrical constraint. This fact is incorporated in the so-called
%FENE model, a particular simple version of which, the FENE-P model
%(e.g. \cite{Chilcotte}),
%is studied in \cite{EH97} and in \cite{AM01}. Other, more elaborate
%versions of the FENE model are currently being developed by \cite{L01}.

A number of different nonlinear constitutive equations which
bound the maximum polymeric stress have been proposed, and
Renardy (\cite{R01,R02}) has considered the asymptotics
of a number of different models.  A key feature of these analyses
is that the thread is predicted to break in finite time when the maximum
elastic stress can no longer balance the capillary pressure
$\gamma/\tilde{h}_{min}$.  For dilute polymer solutions in particular, the
nonlinear form of the force-extension curve close to full extension
is well-established both experimentally and theoretically
(\cite{LLS00, ESGS03}).   Analysis of extensional flow of finitely
extensible nonlinear elastic (FENE) dumbbells results in a maximum
(dimensionless) polymeric stress of order
$\sigma_{zz,max} \approx (2De) G b v' $
where $b = 3N_K$ is the finite
extensibility parameter, which is proportional to the number of Kuhn
steps or links ($N_K$) in the polymer chain. In the elastocapillary
necking regime we obtain $(2De) v' =2/3$ (cf. Figure
\ref{thread}); a naive balance thus suggests that
we require $Gb > 2/3$ for elastic effects to be
able to grow sufficiently large to balance capillary stresses. A more
rigorous balance is substantially more complex and requires a
consideration of the initial polymeric stress in the filament and the
additional strain accumulated by the fluid in the transient process of
the polymeric  stress growing to the saturation value $\sigma_{zz,max}$.

Other, more elaborate  versions of the FENE model have also been
developed (\cite{L01, Ghosh02}) which attempt to capture additional
features of the internal dynamics of the rapid stretching process for long
flexible chains. As the concentration of dissolved polymer
chains is increased, entanglement effects also dramatically modify the
extensional rheology of the viscoelastic fluid. However, recent
experiments have shown that an exponential period of capillary-induced
thinning, followed by finite time breakup still occurs (\cite{PK03}). A
significant benefit of the present experimental configuration and
accompanying analytical description is precisely that the characteristics
of the final breakup process sensitively depend on the nonlinear
description of the test fluid that is used.  Thus analysis of the
capillary-thinning and breakup of polymer solutions provides for a unique
testing ground to better understand some of the important nonlinear
features of viscoelastic constitutive equations at large strains.

%% file: appendix.tex
\appendix
\section{Bounds on the velocity}

In this appendix we estimate the size of the velocity $v(z_0,t)$ 
at some reference point $z_0$ in the crossover region (b) 
(cf. Figure \ref{scheme}); to this end we integrate (\ref{b1})
from the origin of the self-similar region to $z_0$:
\begin{equation}
v(z_0,t)= \frac{v(0,t)h^{2}(0,t) - \int_{0}^{z_{0}}
\left(h^{2}\right)_t dz}{h^2(z_0,t)}
\label{vintegral}
\end{equation}
Using the known solution (\ref{gequ}) one can estimate that the 
main contribution to the integral 
\[
I\equiv\int_{0}^{z_{0}}\left(  h^{2}\right)  _{t}dz
\]
comes from the region where $h(z,t)$ is of order $\sqrt{\ell}$,
$\ell\equiv\exp[-t/(3De)]$ as usual. We therefore introduce the 
rescaled profile $\phi(\overline{z})$ by putting
\begin{equation}
h(z,t) = \sqrt{\ell}\phi(\overline{z}),
\label{defphi}
\end{equation}
for which (\ref{gequ}) simplifies to 
\begin{equation}
\phi^{\prime}\approx-\frac{2\phi}{C_{1}+2\phi^{2}}\label{phi}
\label{phiequ}
\end{equation}
to leading order in $\ell$. Using (\ref{defphi}) one can compute 
the time derivative of $h$ to write the integral $I$ as
\[
I\approx \frac{\ell^2}{3De}\int_{0}^{z_{0}/\ell}\left(-\phi
^{2}+\overline{z}\left(  \phi^{2}\right)  _{\overline{z}}\right)
d\overline{z}.
\]
Integration by parts and a change of variables from $\overline{z}$ to $\phi$
using (\ref{phi}) yield
\[
I\approx\frac{\ell}{3De}z_{0}\phi^{2}(z_{0}/\ell) - 
\frac{2\ell^2}{3De}\int_{0}^{z_{0}/\ell}\phi^{2}
d\overline{z}
\]
\[
\simeq-\frac{2}{3De}z_{0}^{2} + \frac{2\ell^2}{3De}\int_{0}
^{\sqrt{2\left|z_{0}\right|/\ell}}\phi^{2}\frac{C_{1}
+2\phi^{2}}{2\phi}d\phi
\]
\[
\simeq-\frac{2}{3De}z_{0}^{2}+\frac{\ell^2}{6De}\left.  \left(
C_{1}\phi^{2}+\phi^{4}\right)  \right|  _{0}^
{\sqrt{2\left|z_{0}\right|/\ell}}\simeq 
\frac{C_1\ell}{3De}\left|z_{0}\right|.
\]
Here we have used that according to (\ref{match4}) 
$\phi(z_{0}/\ell)\simeq \sqrt{2\left|z_{0}\right|/\ell}$ in the crossover 
region (b). For simplicity, we also assume that the drop radius 
$R=1$. 

Having estimated the integral we find from (\ref{vintegral}),
using that $h(0,t)\simeq h_0\ell$ in the neck region,
\begin{equation}
v(z_{0},t) \simeq \frac{\frac{L}{3De}h_{0}^{2}\ell^2
-\frac{C_1\ell}{3De}\left|z_{0}\right|}{2\left|z_{0}\right|}
\simeq -\frac{C_1\ell}{6De}   \label{ww1}
\end{equation}
as $t\longrightarrow\infty$. From this, the derivatives of $v$
are easily estimated as
\begin{equation}
\frac{\partial v}{\partial t}(z_{0},t)\simeq \frac{\ell}{18De^2}
\label{ww2}
\end{equation}
and from (\ref{b1}) one finds
\begin{equation}
v_{z}(z_{0},t)=\frac{-v(z_{0},t)h_{z}(z_{0},t)-h_{t}(z_{0},t)}{h(z_{0},t)}
\simeq \frac{\ell}{6De}{\left|  z_{0}\right|  }\label{ww3}
\end{equation}

Finally, we use (\ref{ww1}), (\ref{ww2}), and (\ref{ww3}) to estimate the 
different terms of the second equation of (\ref{fse}). While for example
$\ell^2\dot{\overline{v}} \simeq \ell^3/De^2$, the curvature term on the
right of (\ref{fse}) can 
be estimated as 
$\left[\overline{h}^2 K\{\overline{h}\}\right]'/\overline{h}^2 \simeq
\ell^{1/2}/\overline{z}^{3/2}$, which is always much larger in the 
transition region (b). Therefore the right hand side of (\ref{fse})
always dominates the left hand side in the transition region.